\documentclass{article}
\usepackage{graphicx} 
\usepackage{subcaption} 
\usepackage{amsmath,amsfonts,amsrefs,amssymb}

\newtheorem{theorem}{Theorem}[section]

\newtheorem{corollary}[theorem]{Corollary}

\def\qed{\hfill\rule{1ex}{1ex}\\}

\DeclareMathOperator{\sgn}{sgn}

\title{Comparison of the potential energy for different equilibrium configurations of symmetric and asymmetric floating drops}
\author{Mason Mault\footnote{Department of Mathematics, Texas State University, 601 University Dr., San Marcos, TX 78666, mdm353@txstate.edu} and Ray Treinen\footnote{Department of Mathematics, Texas State University, 601 University Dr., San Marcos, TX 78666, rt30@txstate.edu}}
\date{\today}

\begin{document}

\maketitle

\begin{abstract}
We provide a numerical method for computing solutions to a free boundary problem arising from the equilibrium state of a floating drop.  This numerical method is based on a Newton's method for the underlying nonlinear boundary value problems, and at each iterative step a Chebyshev spectral collocation method is employed.  The problems considered here are those that can be described by using generating curves, and include problems in $\mathbb{R}^2$ and $\mathbb{R}^3$.  

The resulting nine-dimensional space of physical parameters is explored, and examples are given that highlight the potential energy of centrally located drops, wall-bound drops, and asymmetrical configurations in $\mathbb{R}^2$.  Non-uniqueness of solutions to the corresponding Euler-Lagrange equations is displayed, and also strong evidence of non-uniqueness of energy minimizers is given.
\end{abstract}

\section{Introduction}
\label{intro}
\begin{figure}[h!]
	\centering
	\includegraphics[scale=.2]{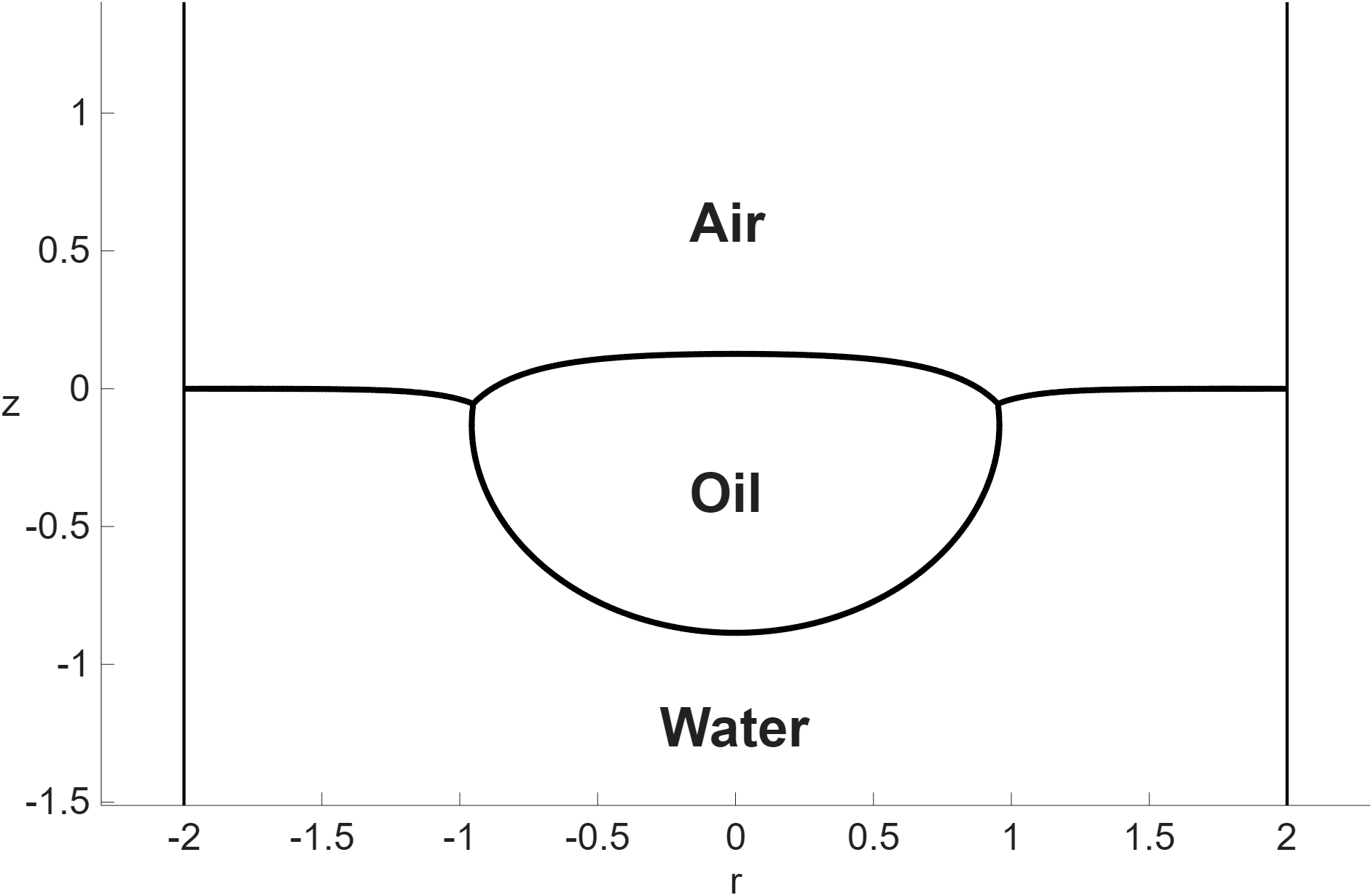}
	\caption{Schematically, a drop of oil floats on the surface of water inside a capillary tube.  Shown is a vertical section of an axially symmetric configuration in $\mathbb{R}^3$.  The capillary tube is indicated by the two vertical lines, and the tube walls will be omitted in all other figures in this work.}
	\label{fig:airOilWater}
\end{figure}

We consider the problem of finding the equilibrium state of three immiscible fluids in a laterally bounded container, sometimes known as a floating drop problem. A guiding example would be water rising in a capillary tube with a small drop of oil introduced to the system, as is shown in Figure~\ref{fig:airOilWater}.  
In every case, we will find numerical solutions for these configurations are always non-unique, and we will also show many examples of symmetry breaking with a lower-energy state given by an asymmetric configuration.  
What is particularly striking is that  some examples are found where the energy minimizer is not unique.

A little more detail is needed to be specific.  Given a laterally bounded container partially filled by a fluid, we introduce a prescribed volume of a different fluid, known as the drop, and we consider here only drops with densities less than the density of the supporting fluid.  As we will see in the next section, there are nine physical parameters of interest in this problem, and we explore this parameter space for examples that highlight the symmetry and asymmetry of the resulting configurations as well as other interesting phenomena.  Our methods are computational, and not analytical, however we will comment on the limited theory as it currently sits. For any selection of physical parameters we find configurations where the drop is adjacent to the wall of the container, and we also find configurations with the drop floating in the center of the container. There are examples of non-uniqueness of solutions to the resulting free boundary problems for every set of parameters we considered.  We compute the potential energy of both configurations of the wall-bound drops and the centrally floating drops to determine which configuration has the lower energy.  Surprisingly, there are many selections of parameters where the two configurations have the same potential energy.    This is the first time non-uniqueness has been observed with this model of the physical configuration, both at the level of the competing floating drop configurations, as well as the configurations that produce energy minimizers.

Our approach to floating drop problems follows the framework established by Elcrat, Neel, and Siegel \cite{ENS}.
Historically, the mathematical study of three immiscible fluids in equilibrium goes back to at least 1806, when Laplace considered mercury on water \cite{Laplace}.
Massari \cite{Massari},  Leonardi \cite{Leonardi}, Elcrat and Treinen \cite{ET2011}, and Blank, Elcrat, and Treinen \cite{BET} studied floating drops using functions of bounded variation.  In this direction, both Maggi \cite{Maggi2012} and Morgan \cite{MorganGMT} consider fluid clusters in their books, with further references therein.  Ickes and Treinen \cite{IckesTreinen} combined floating drops with additional floating rigid objects.  Slobozhanin \cite{slobozhanin} contributed to the problem, and also see Gibbs \cite{Gibbs} as a precursor for \cite{ENS}. Treinen \cite{Treinen2010} gave a general existence theorem for symmetric configurations,  then he established the symmetry of some configurations \cite{Treinen2012b}, and also considered some results for the unbounded problem \cite{Treinen2023b}. Lawlor and Morgan \cite{LawlorMorgan}, Morgan \cite{Morgan}, and White \cite{White} treat three immiscible fluids using more general geometric measure theory.   Numerical simulations were also done by Elcrat and Treinen \cite{ET2005}, and that work can be seen as the approach we  both improve upon and extend to new configurations here.

The theory at this point is largely of two types.  The first type is primarily developed in general terms of the existence of energy minimizers and associated properties.  The second type gives precise geometric information in the form of solutions to differential equations and is largely based on symmetric configurations in an unbounded domain.  The paper by Treinen \cite{Treinen2010} takes this approach and gives the existence of solutions to the associated free boundary problem for many radially symmetric configurations, but this is given in terms of properties of component interfaces, not all of which are completed.  The three dimensional problems we will explore in this work fit into this framework, and the symmetric drops that are adjacent to the tube wall are completely new.  Also, there is no complete proof of symmetry in any of the floating drop problems.  Treinen \cite{Treinen2012b} has a symmetry result for the unbounded problem, but the theorem has a hypothesis that has not been shown to be satisfied.  This hypothesis is related to other geometric problems involving the mean curvature operator under consideration here, and is currently the best result available.  As we shall see in this work, that unbounded configuration is the only venue where such a theorem can be stated in generality.  Finally, Elcrat and Treinen \cite{ET2005} computed a centrally located drop in 3D, but none of the other configurations we will consider appear in that work.

Physically, the appropriate approach is to study two-dimensional surfaces describing the boundaries between the fluids, and the fluids are described as subsets of $\mathbb{R}^3$.  We restrict our attention to those problems described by ordinary differential equations.  For our first consideration this means that we are restricting our interfaces to be radially symmetric about a vertical axis.  It would also be interesting to study the full three-dimensional problem without this symmetry, however that would require techniques beyond the scope of this project.

We will also consider the lower-dimensional model in $\mathbb{R}^2$.  The arguments in \cite{ENS} that are used here do not depend on the underlying dimension of the problem.    As we will see, our solutions also satisfy some ordinary differential equations, but we will be able to identify examples of  symmetry breaking in this case.
This type of configuration has a physical interpretation beyond the lower-dimensional analogue of the problem.  If one considers two vertical plates at $y = \pm \epsilon$ for a small enough $\epsilon > 0$ and  with homogeneous wetting energies on these vertical plates so that the contact angles are all $\pi/2$ there, then the floating drop problem will be a horizontal cylinder traced out by the lower-dimensional problem.    Of course, if $\epsilon$ is allowed to be large, then the stability of these drops becomes unlikely.   From either perspective, we will treat this lower-dimensional setting, and we will replace full dimensional quantities of volume and surface area with the corresponding quantities of area and length as appropriate.

There is a history of studying capillary surfaces in this lower dimensional setting, with examples given in Aspley, He, and McCuan \cite{AspleyHeMcCuan2015}, McCuan \cite{McCuan2013}, \cite{McCuan2015}, \cite{McCuan2017}, \cite{McCuan2022}, McCuan and Treinen \cite{McCuanTreinen2018}, and Wente \cite{Wente2006}.  The related Hele-Shaw problem is classically derived for a two dimensional setting even though many applications are three dimensional.  Recently, a study of a three-phase Hele-Shaw model appeared by Zhao, Barua, Lowengrub, Ying, and Li \cite{ZhaoBaruaLowengrubYingLi2024}.  We also mention the long history of the use of complex variables in the study of fluid mechanics, which is fundamentally two dimensional.

In the next section the mathematical model and numerical methods are discussed, concluding with a discussion of the free parameters in these problems.  Section~\ref{sec:3D} collects our experiments in $\mathbb{R}^3$.   Specifically, in Subsection~\ref{sec:physical} an example experiment is given where the volume of a drop is increased, and the resulting energy profiles of centrally located drops and wall-bound drops are compared.  In Subsection~\ref{sec:3dmore} other examples in $\mathbb{R}^3$ are explored, following the general framework established for the different types of parameters.  A heuristic is given for when a centrally located drop might tend to be the energy minimizer.  In Section~\ref{sec:2d} floating drops in $\mathbb{R}^2$ are considered, with Subsection~\ref{sec:2Dmore} collecting more overviews of parameter studies, and an example is given where the heuristic guide provided in Subsection~\ref{sec:3dmore} is false.  Then in Subsection~\ref{sec:asymmetry} the symmetry of energy minimizers is explored, and some asymmetrical configurations are highlighted.   In Section~\ref{Zero} we treat a zero-gravity case and use that to validate our numerical results.  Finally, in Section~\ref{sec:conclusions} we summarize our conclusions.

Codes to produce the floating drops considered in this work can be found at \\
\texttt{https://github.com/raytreinen/Floating-Drops}

We would like to thank Nestor Guillen for his generous support, John McCuan for interesting conversations on floating drops, and the anonymous referees for their valuable comments. The first author was partially supported by the NSF Grant, DMS-2144232.

\section{The mathematical model and basic algorithms}
\label{model}

In this section we introduce the mathematical model.  As the parameters involved in our approach  naturally lead to a numerical algorithm, we discuss our numerical approach in the same exposition.  This framework originated in \cite{ENS} and was elaborated on by Treinen \cite{Treinen2010}.

\subsection{Energy and related parameters}

We consider the interior of the capillary tube as a domain $\Omega$, which is either as a subset of $\mathbb{R}^2$ or $\mathbb{R}^3$ that is laterally bounded, and $\partial\Omega$ extends vertically as an infinite cylinder or two vertical lines if $\Omega\subset\mathbb{R}^2$.  Specifically, $\Omega = B \times \mathbb{R}$ for a ball $B$ in $\mathbb{R}$ or $\mathbb{R}^2$.  
For the rest of this section, we use terms volume and surface area as applied to the model in $\mathbb{R}^3$, and we make comments about the lower dimensional setting as needed.
There are three fluids, represented by the sets $E_0,E_1,E_2$ that must satisfy $\bar E_0\cup\bar E_1 \cup \bar E_2 = \bar\Omega$.  The set $E_1$ represents the drop and has a finite volume, while the other two fluids have infinite volume.  The boundaries of the fluids are assumed to be $C^2$ surfaces away from the triple junction, $\Gamma$, which we assume is a smooth curve in $\mathbb{R}^3$, or either one or two triple-junction points in $\mathbb{R}^2$, and we assume that the interfaces between any two of the fluids is $C^1$ up to $\Gamma$ and $\partial \Omega$ as appropriate.  These surfaces are not assumed to be graphs over $B$ in general.  

The interface between $E_i$ and $E_j$ is denoted by $S_{ij}$, where $S_{ij} = S_{ji}$.
When convenient, we will use function notation for these surfaces, and when we do so, we fix $u$ for $S_{01}$, $v$ for $S_{12}$, and $w$ for $S_{02}$.  Associated to each surface is a surface tension: $\sigma_{01},\sigma_{12},\sigma_{02}$, giving a surface energy that can be written as the surface tension times the area of each surface.
The force balance condition
\begin{eqnarray}
\sigma_{01} &\leq& \sigma_{12} + \sigma_{02}, \label{fb1}\nonumber\\
\sigma_{12} &\leq& \sigma_{01} + \sigma_{02}, \label{fb2}\\
\sigma_{02} &\leq& \sigma_{01} + \sigma_{12} \label{fb3} \nonumber
\end{eqnarray}
is assumed.   This is necessary for equilibrium, as described in \cite{BET},  \cite{ENS}, \cite{ET2011}, and \cite{Massari}. 

\begin{figure}[h]
	\centering
	\includegraphics[scale=.3]{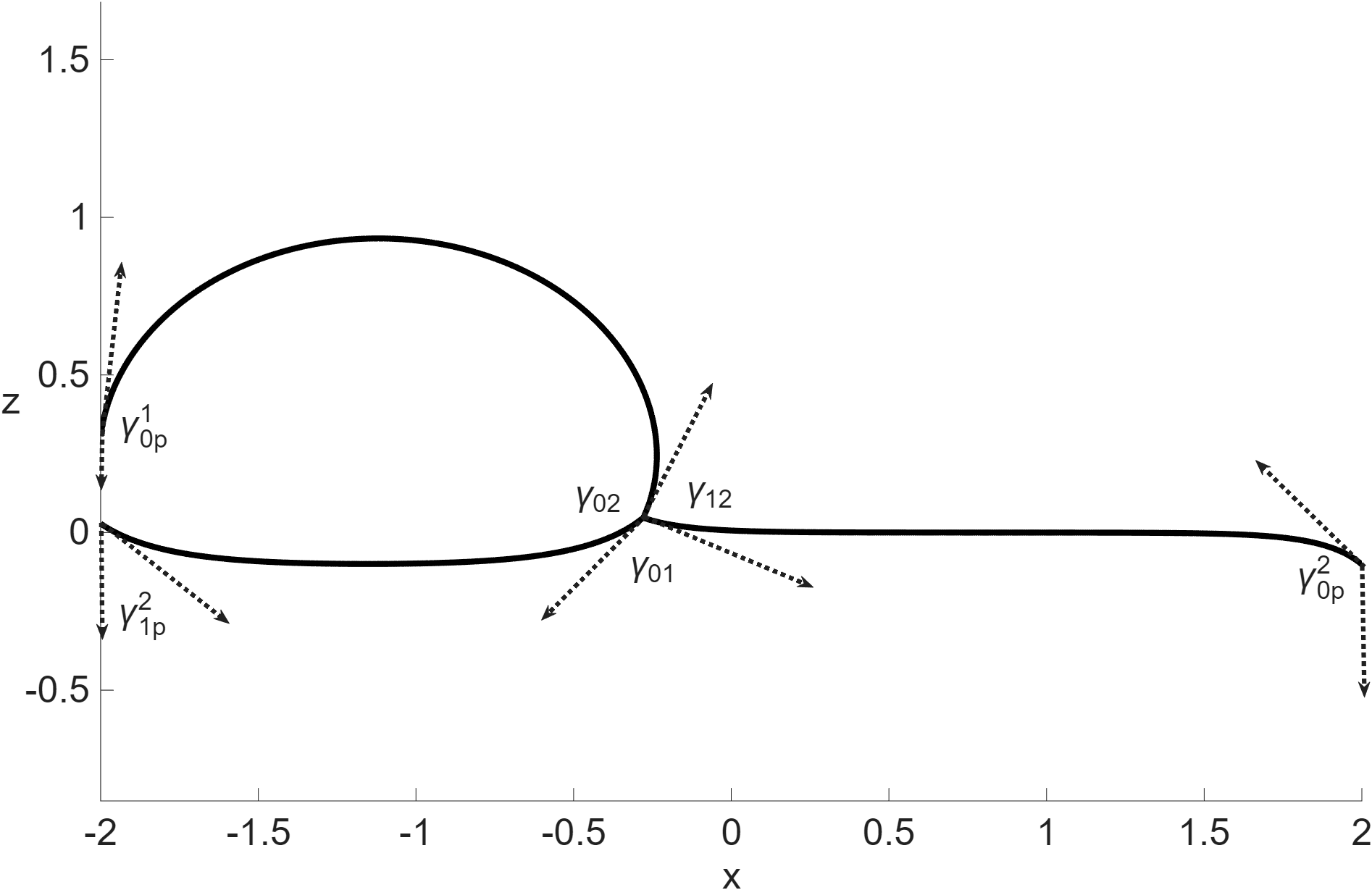}
	\caption{The surface tensions are shown in a force balance form as vectors, and the associated interior contact angles follow.  The plate contact angles are also shown on the walls of the tube.  The floating drop pictured here is a wall-bound drop in $\mathbb{R}^2$.}
	\label{fig:contactAngles}
\end{figure}

The contact angle between $S_{ij}$ and $S_{ik}$ is denoted by $\gamma_{jk}$.
The contact angles and surface tensions must satisfy the Neumann triangle relation:
\begin{equation}\label{neumann_triangle}
\frac{\sin\gamma_{01}}{\sigma_{01}} = \frac{\sin\gamma_{12}}{\sigma_{12}} = \frac{\sin\gamma_{02}}{\sigma_{02}},
\end{equation}
as proved in \cite{ENS}, and this is illustrated in Figure~\ref{fig:contactAngles}.  Each fluid $E_j$ has an associated density $\rho_j$ and we only consider the case $\rho_0\leq\rho_1\leq\rho_2$.  Then, for each fluid, the potential energy due to gravity is generically
$$
\int_E \rho g z \, dV
$$
with horizontal slices taken at each height $z$.  We can iterate that integral so that we have $\int(\int_0^u z\, dz) = 0.5\int (\sgn u) u^2$ for a generic surface with height $u$, though the domain of the second integral depends on the geometry of the configuration and we will say more about this later.  For now we note that this allows us to consider finite quantities for the gravitational potential energy in reference to a height $z = 0$. This reference level is based on the capillary tube being placed in an exterior bath of liquid $E_2$, and that bath of liquid is unbounded, with the reference height for its interface given by a limiting height as the distance to the tube goes to infinity.  Of course, we take a coordinate system so that this reference height is zero.
Each surface has an associated capillary constant given by  
\begin{equation}
\kappa_{01} = g\frac{\rho_1 - \rho_0}{\sigma_{01}},\quad \kappa_{12}= g\frac{\rho_2 - \rho_1}{\sigma_{12}}, \quad \kappa_{02} = g\frac{\rho_2 - \rho_0}{\sigma_{02}},
\end{equation}
and we normalize this with the assumption that $\rho_0 = 0$.

The final component is a wetting energy, and this quantity drives the fluid rise of a meniscus in a tube.  Here we borrow our terminology from the sessile drop technique for measuring contact angles where a drop of fluid is placed on a horizontal plate.  Interestingly, this technique goes back to Adams and Bashforth and their multi-step method \cite{BashforthAdams}.  In that classical sessile drop problem, a single drop volume rests on a horizontal plate.  The contact angle is measured within the drop at the intersection of the fluid interface and the plate.  In our configurations there will be three such possible contact angles.  If the drop is floating in the center of the tube, then the only contact angle that is present is that of the bulk fluid in the tube.  If, however, the drop itself is touching the tube wall, then the upper and lower interfaces of the drop each have their own wetting contact angle.

If we take a convention on labeling the contact angles at the wall by using the super-script to indicate the set on whose interior the angle is measured, and the subscript contains the labels showing the adjacent fluid and that this is a ``plate'' angle.   The wetting energy is a constant multiplied by the surface area of the tube in contact with a given fluid, but since that is infinite for two of the liquids in our model, we simply use the reference height of zero so that wetted areas corresponding to negative heights give negative surface area. 
Then, using the variational argument found in Finn \cite{ecs}, or more pertinent, his paper \cite{Finn2006}, Elcrat, Neel, and Siegel \cite{ENS}, Maggi \cite{Maggi2012}, or McCuan and Treinen \cite{McCuanTreinen2013}, we can relate $\cos\gamma^2_{1p}$ to the other angles. Take note that the wetting coefficient of $E_1$ is $-\sigma_{01}\cos\gamma^1_{0p}$ and the wetting coefficient of $E_2$ is $-\sigma_{02}\cos\gamma^2_{0p}$, and then, using  Finn's derivation \cite{Finn2006},
\begin{equation}
\label{eqn:plate}
\sigma_{12}\cos\gamma^2_{1p} = -\sigma_{01}\cos\gamma^1_{0p} + \sigma_{02}\cos\gamma^2_{0p}.
\end{equation}
We will use two of these plate angles to derive the third, and we choose a particular pair depending on what is motivating the experiment at the time.  These plate contact angles are also depicted in Figure~\ref{fig:contactAngles}.  For a given physical configuration the properties of the fluids including the wetting are treated as fixed and known.  Any two terms in \eqref{eqn:plate} determine the third even in the case where only one of the terms appears in the energy because the drop does not wet the wall in that configuration.  In that sense we are able to compare with other configurations where the previously fixed , but unused physical constants are now relevant.

Floating drops may be seen as critical points or (local) minimizers of the energy functional
\begin{eqnarray}
\mathcal{E} &:=&  \sigma_{01}|S_{01}| + \sigma_{02}|S_{02}| + \sigma_{12}|S_{12}| + \sum_{i=0}^2\rho_i g \int_{E_i} z\, dV  \nonumber\\
& & - \sigma_{01}\cos\gamma^1_{0p}|\partial E_1 \cap \partial \Omega| - \sigma_{02}\cos\gamma^2_{0p}|\partial E_2 \cap \partial \Omega|.
\end{eqnarray}

We state this quantity in the framework of the full dimensional problem, with the first three terms collecting the surface area quantities, the sum in the fourth term containing a volume integral, and the last two terms containing the wetted areas on the tube wall. We can easily adapt this to the lower dimensional case by considering the interface lengths in the first three terms, and modifying the integrals from volume integrals to area integrals, and the wetting terms then have wetted lengths on the tube walls, which are two vertical lines in the vertical plane.

\subsection{Boundary value problems and a Chebyshev  spectral collocation method}

It was shown in \cite{ENS} by a variational argument that the interfaces satisfy the following Young-Laplace equations and free boundary condition:
\begin{eqnarray}
2H_{S_{01}}(x,y,z) &=& \kappa_{01} z + \lambda/\sigma_{01}, \label{equ}\\
2H_{S_{12}}(x,y,z) &=& \kappa_{12} z - \lambda/\sigma_{12} , \label{eqv}\\
2H_{S_{02}}(x,y,z)  &=& \kappa_{02} z, \label{eqw}
\end{eqnarray}
\vspace{-2.5ex}
\begin{equation}\label{eqnt}
S_{01},S_{12},S_{02} \mbox{ meet at $\Gamma$ and satisfy \eqref{neumann_triangle}}.
\end{equation}
Here $H_S(x,y,z)$ is defined to be the mean curvature of the surface $S$ when $(x,y,z)\in S$ and is undefined otherwise.  Often $z=z(x,y)$, then $2H_S(x,y,z) = \nabla \cdot Tz$ with $Tz = \nabla z /\sqrt{1 + |\nabla z|^2}$, though we will not assume this non-parametric form here.  We also note that Treinen \cite{Treinen2010} adapted this theoretical approach to some of the configurations in our current work. From this point onward, we will use the convention of capitalizing the height of the solutions of the differential equations without a Lagrange multiplier, and lower case letters will be reserved for the height of the physical interfaces.

Two particular forms of the differential equation $2H_U = \kappa U$ will sometimes be used.  The first form assumes the solution in $\mathbb{R}^3$ is symmetric about the vertical axis, parametrized by arc length $s$, and is
\begin{eqnarray}
\frac{dr}{ds} &=& \cos\psi, \label{eqn:drds}\\
\frac{dU}{ds} &=& \sin\psi, \label{eqn:duds}\\
\frac{d\psi}{ds} &=& \kappa U - \frac{\sin\psi}{r}.\label{eqn:dpsids}
\end{eqnarray}
Here $r$ is the radial coordinate of the solution, $U$ is the height, and $\psi$ is the inclination angle measured from the positive radial axis.  This form allows for inflection points and vertical points, where they exist. The next form is for solutions in $\mathbb{R}^2$, and is similar:
\begin{eqnarray}
\frac{dx}{ds} &=& \cos\psi, \label{eqn:dxds}\\
\frac{dU}{ds} &=& \sin\psi, \label{eqn:duds2}\\
\frac{d\psi}{ds} &=& \kappa U,\label{eqn:dpsids2}
\end{eqnarray}
where the horizontal component of interface graph is $x$.


The appropriate boundary conditions for these systems of ODEs are to prescribe the inclination angle $\psi = \psi_b\in[-\pi,\pi]$ at some radius $r = b$ (or $x = b$).  If the surface spans the vertical axis, then we prescribe $\psi = -\psi_b$ at $r = -b$, and these interfaces are well studied, with Finn \cite{ecs} being an excellent reference for the system in $\mathbb{R}^3$.  If instead the surface is topologically annular, then we alternatively prescribe $\psi = \psi_a\in[-\pi,\pi]$ and $r = a$ for some $0 < a < b$.  These annular capillary surfaces are not as well understood.  In $\mathbb{R}^3$, the papers by Elcrat, Kim, and Treinen \cite{EKT}, Gordon and Siegel \cite{GordonSiegel2010a}, \cite{GordonSiegel2010b}, Siegel \cite{Siegel2006}, and Treinen \cite{Treinen2012a} collect the theory as we know it, and Bagley and Treinen \cite{BagleyTreinen2018} present a united approach.  It should be noted that this theory is not yet rich enough to establish the same conclusions that appear in \cite{ENS} where they used an unbounded liquid bridge.  In $\mathbb{R}^2$ the surface (curve) is independent of horizontal translation, and is only fixed by the difference between $a$ and $b$.  The papers by McCuan \cite{McCuan2015}, \cite{McCuan2022}, McCuan and Treinen \cite{McCuanTreinen2018}, and Wente \cite{Wente2006} collect at least most of the relevant theory.

The basic idea is to use the systems of differential equations without Lagrange multipliers and solve boundary value problems over some domain with free boundary  $\Gamma$ and appropriate physical conditions for $\psi$ there so that the interfaces $U$ and $V$ that correspond to $S_{01}$ and $S_{12}$ are (generally) positive.

We use the Chebyshev  spectral collocation methods developed for these equations by Treinen \cite{Treinen2023a}.  The methods in that paper follow the rectangular collocation techniques of Driscoll and Hale \cite{DriscollHale2016} and the general overview by Aurentz and Trefethen \cite{AurentzTrefethen2017}.  Also see Trefethen \cite{Trefethen2000}.  These methods use a Newton method to treat the non-linearity of the equations, Chebyshev spectral methods to solve the resulting iterated linear systems, a scaling trick to compute the total arc-length of the solution of the boundary value problem as part of the system, and a continuity method for solutions that are not a graph over the base domain.  The algorithm is adaptive and the tolerances are set to be within fourteen digits of relative error for the Newton step and ten digits of the relative error for the boundary value problem.  We do not further repeat any discussions of the error and refer the reader to Treinen \cite{Treinen2023a}.  We note that this level of accuracy is likely more accurate than the underlying mathematical model given that the width of a water molecule is about 2.75 angstroms, which is four orders of magnitude greater than the error tolerance we are specifying if we take our units in the MKS system.  

The programs used will then generate data points generically along $(r,U,\psi)$ associated with a number of Chebyshev points on an interval of length $2\ell$, which is the total arc-length of the generating curve.  This is typically achieved with $n + 1 = 14$ Chebyshev points.  We again refer the reader to Treinen \cite{Treinen2023a} for more details.  This approach is robust and is a significant improvement over the sometimes unstable shooting methods used by Elcrat and Treinen \cite{ET2005}, though it should be pointed out that there are extreme cases where the algorithm used here could fail to converge, even when modified to localize problematic portions of the solution curve using domain decomposition, as is discussed by Haug and Treinen \cite{HaugTreinen2024}.  In our experiments for this work we have called this solver  millions of times and have never encountered the degenerate cases that were explored by Haug and Treinen.  From this point onward, when we mention solutions to a particular boundary value problem, we mean a numerical approximation that is found with the algorithm from Treinen \cite{Treinen2023a} using the tolerances we have mentioned above.

\begin{figure}[t]
	\centering
	\includegraphics[scale=.2]{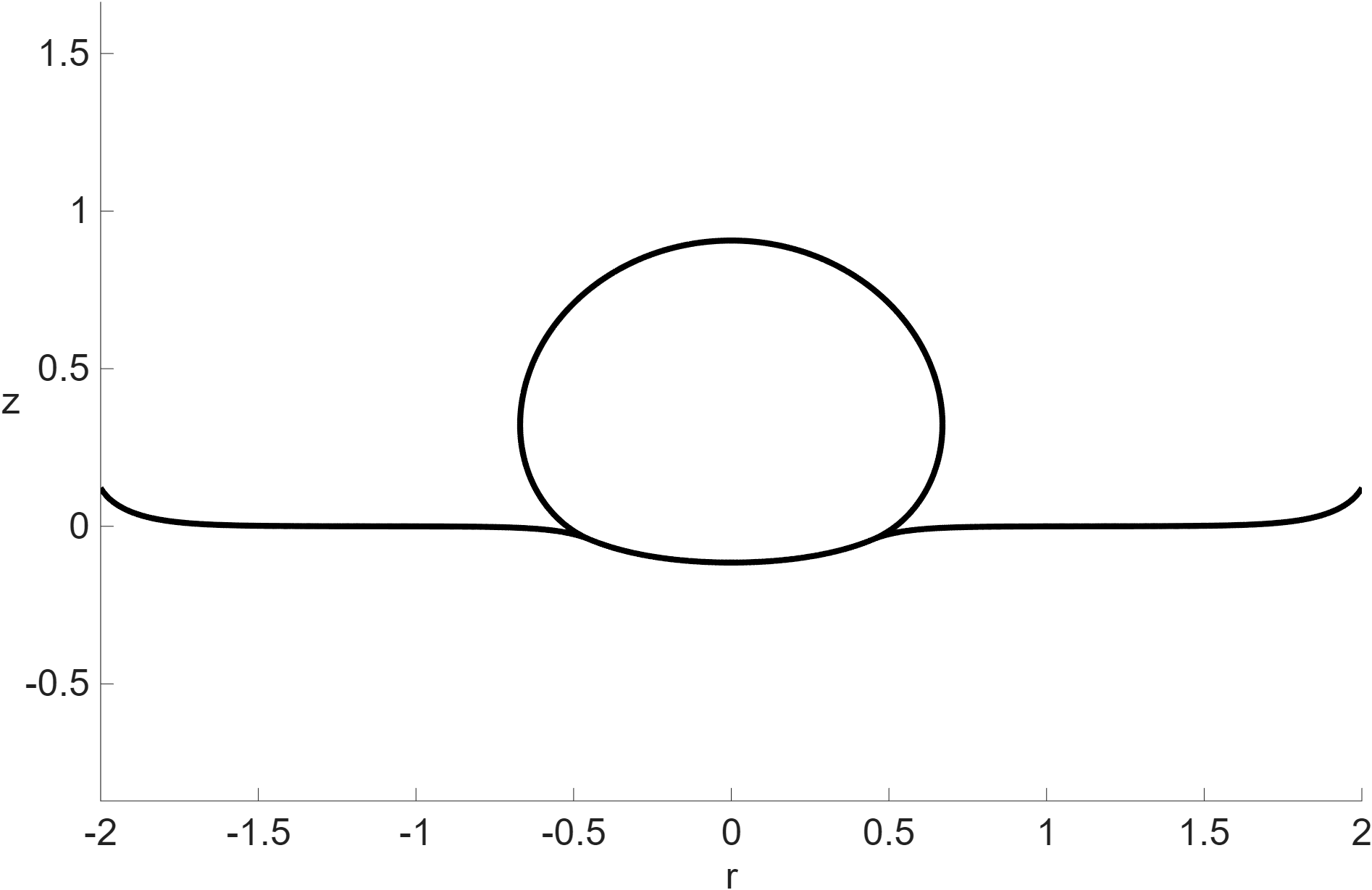}
	\caption{A typical centrally located floating drop.  The physical parameters that determine this drop are a volume of 1, tube radius of $R=2$, densities $\rho_1 = 1$ and $\rho_2 = 15$, surface tensions $\sigma_{01} = 6, \sigma_{02} = 2.0001$, and $\sigma_{12} = 7.9999$, and plate angle $\gamma^2_{0p} = 0.5$.}
	\label{fig:3dCentralDrop1}
\end{figure}
\subsection{Matching physical surfaces}

Next, we note that in \cite{ENS} the Lagrange multiplier $\lambda$ was eliminated by matching the interfaces at the free boundary $\Gamma$, which is at a particular radius $r = \bar r$, and they found
\begin{equation}
    \label{Eq: lagrange multiplier}
    \lambda = (\bar{U} + \bar{V})\frac{\kappa_{01}\sigma_{01}\kappa_{12}\sigma_{12}}{\kappa_{01}\sigma_{01} + \kappa_{12}\sigma_{12}}
\end{equation}
with $\bar{U}$ and $\bar{V}$ representing the heights at $\bar r$.  The same formula holds for the lower dimensional model with $x = \bar x$.  
Then eliminating $\lambda$ results in physical interfaces
\begin{eqnarray}
    \label{Eq: lambda shift u}
    u = \frac{\lambda}{\kappa_{01}\sigma_{01}} - U \\
    \label{Eq: lambda shift v}
    v = V - \frac{\lambda}{\kappa_{12}\sigma_{12}}
\end{eqnarray}
that match at $\bar r$ (or $\bar x$).   With the surfaces $u$ and $v$ matched at some fixed $\bar r > 0$ (or $\bar x > 0$), we do not in general have $w = u = v$ there.

Before proceeding to match $w = u = v$ at $\bar r$, we need to describe the geometry of the interfaces near the free boundary.
To fix the concepts, consider first a centrally located drop that is symmetric about the vertical axis, as shown in Figure~\ref{fig:3dCentralDrop1}.  Then the boundary conditions on $v$ depend on the inclination of $v$ at $\bar r$.  We denote this crucial quantity by $\bar\psi$,  and the boundary conditions for $v$ are then that the inclination angle of that surface is $-\bar\psi$ at $r = -\bar r$ and $\bar\psi$ at $r = \bar r$.
Then we can use \eqref{neumann_triangle} and the surface tension values to formulate boundary conditions for $u$ based on $\bar\psi$.  Figure~\ref{fig:contactAngles} is a useful companion for deriving these boundary conditions.
It follows that $U$ has inclination angles $-(\gamma_{02} - \bar \psi)$ at $-\bar r$ and $\gamma_{02} - \bar \psi$ at $\bar r$.
We also have an inclination angle of $\bar\psi + \gamma_{01} - \pi$ for $w$ at $\bar r$, and $w$ satisfies $\sin\psi = \cos\gamma^2_{0p}$ at $r = R$.

Then we note that $\bar\psi = 0$ gives $V \equiv 0$ and $v < 0$ while $w$ is positive at $\bar r$, which can be proved in many cases, but is in general merely an observed fact.  The conditions on these component interfaces were discussed by Treinen \cite{Treinen2010}.  Then when $\bar\psi = \gamma_{02}$, $U\equiv 0$ and $u > 0$, while $w$ is (observed to be) negative at $\bar r$.  Thus defining $\bar v(\bar\psi)$ and $\bar w(\bar\psi)$ to be the heights of $v$ and $w$ at $\bar r$ for a particular $\bar\psi$, we can then define
\begin{equation}
F(\bar\psi) := \bar v(\bar\psi) - \bar w(\bar\psi)
\end{equation}
and we observe that there is some value of $\bar\psi \in (0,\gamma_{02})$ that is a zero of $F$.  We find this value of $\bar\psi$ numerically using Matlab's fzero function applied to $F(\bar\psi)$ with an initial guess of $\bar\psi = \gamma_{02}/2$.  The tolerance of Matlab's fzero algorithm defaults to near machine precision.  The repeated calls to $F(\bar\psi)$ will generate surfaces $u, v$, and $w$ for each attempted value of $\bar\psi$, so the spectral solver is used multiple times inside this fzero process.

Then with this critical value of $\bar\psi$, we have matched $u$, $v$, and $w$ at a prescribed $\bar r> 0$.  

\subsection{Drop volumes}

In general, the enclosed volume of the drop just formed will not match the prescribed volume of the drop.   What remains in the floating drop problem is to match the prescribed volume. 

In order to efficiently compute the volume, we derive two volume formulas, depending on whether the drop is centrally located or bound to the wall.  First we find an identity for a generic $(r,V,\psi)$ solution to \eqref{eqn:drds}-\eqref{eqn:dpsids}:
\begin{eqnarray}\label{eqn:identity}
\frac{d}{ds} \left( r\sin\psi \right) &=& \cos\psi\sin\psi + r\cos\psi\left(  \kappa u V - \frac{\sin\psi}{r} \right) \nonumber\\
&=& \kappa r V\cos\psi.
\end{eqnarray}
Then the centrally located drop is made up of two volumes of fluid, divided by the horizontal plane $z\equiv \bar v$.  We will compute the two component volumes $\mbox{Vol}_U$ and $\mbox{Vol}_V$ using the standard representation, which is merely a rigid motion of the physical component volumes.  Then
\begin{eqnarray}\label{eqn:3DCentralVolume}
\mbox{Vol}_V &=& 2\pi\int^{\bar r}_0 \left( \bar V - V  \right) r\, dr \nonumber\\
&=& \pi\bar V {\bar r}^2 - \int^\ell_0 \rho V \frac{dr}{ds}\, ds \nonumber\\
&=& \pi\bar V {\bar r}^2 - \frac{2\pi}{\kappa_{12}}\int^\ell_0 \frac{d}{ds}\left( r \sin\psi \right)\, ds \nonumber\\
&=& \pi\bar V {\bar r}^2 - \frac{2\pi}{\kappa_{12}} \bar r \sin\bar\psi
\end{eqnarray}
and similarly 
\begin{equation}
\mbox{Vol}_U = \pi\bar U {\bar r}^2 - \frac{2\pi}{\kappa_{01}} \bar r \sin\left(  \gamma_{02} - \bar\psi \right).
\end{equation}
We obtain the total volume of $\mbox{Vol} = \mbox{Vol}_U + \mbox{Vol}_V$
for the centrally located drop.
We have chosen to present a cleaner partial derivation that assumes the interfaces do not continue past $\bar \psi = \pi/2$, though the same formula holds for those larger $\bar \psi$ values as well.  We also note that if either of the interfaces $U$ or $V$ passes above the corresponding $\bar U$ or $\bar V$, then that portion of the volume will be negative.  However, the other interface will be adding exactly that missing volume to the total.  We will also exclude any non-physical drops where $u$ and $v$ intersect away from $\bar r$.

Then we have a second fzero module where we match the computed volume to the prescribed volume by varying $\bar r$.  The initial guess for $\bar r$ is generally $0.2\sqrt[3]{\mbox{Vol}}$, but the factor of $0.2$ may need to be adjusted in some cases.  In some of the automation that will follow in this paper, this factor may need to be changed adaptively.  The output of these two nested loops gives a solution to the floating drop problem with a centrally located drop.  


\begin{figure}[t]
	\centering
	\includegraphics[scale=.2]{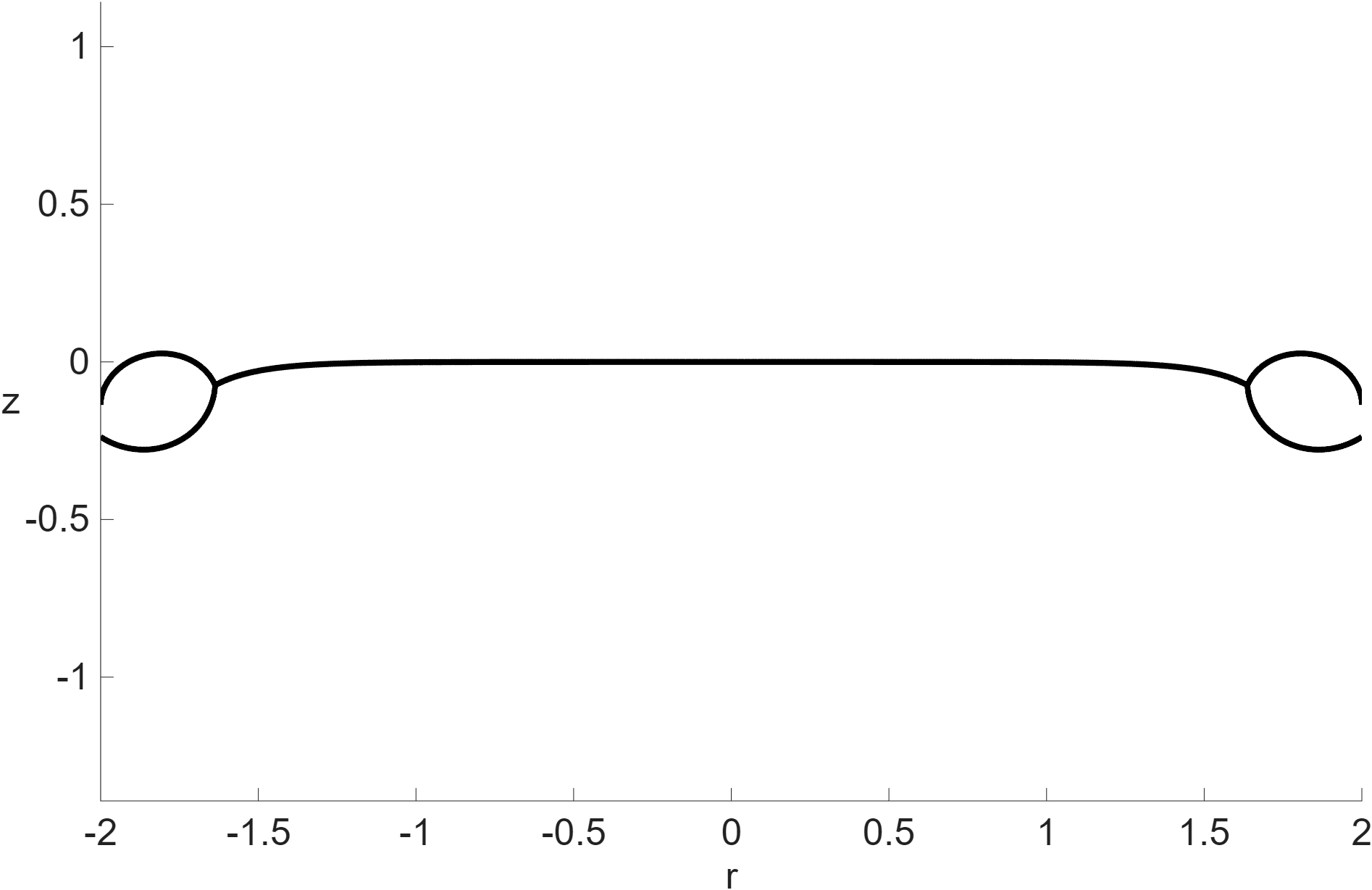}
	\caption{A typical wall-bound floating drop.  The physical parameters that determine this drop are a volume of 1, tube radius of $R=2$, densities $\rho_1 = 14$ and $\rho_2 = 15$, surface tensions $\sigma_{01} = 6.01, \sigma_{02} = 2.99$, and $\sigma_{12} = 7$, and plate angles $\gamma^1_{0p} = \pi - 0.01$ and $\gamma^2_{1p} = 1$.}
	\label{fig:3dWallDrop1}
\end{figure}

To compute radially symmetric wall-bound drops in $\mathbb{R}^3$,  recall that the drop forms a ring around the edge of the wall, and we work with a section of this radially symmetric configuration, as is shown in Figure~\ref{fig:3dWallDrop1}.  We adapt the above process to the left side of the configuration, where the generating curves start at radius $-R$ and the upper and lower portions of the drop meet at $-\bar r$.  
Starting with the lower interface $V$, this interface has inclination angles $\gamma^2_{1p} - \frac{\pi}{2}$ at $r = -R$ and  $\bar\psi$ at $r = -\bar r$  with $\bar\psi\in(0,\gamma_{02})$.   
Then $U$ has inclination angle $\frac{\pi}{2} - \gamma^1_{0p}$ at $r = -R$ and $\gamma_{02} - \bar \psi$ at $r = -\bar r$.   Finally, $w$ has inclination angle $\bar\psi + \gamma_{01} - \pi$ at $r = -\bar r$ and $-(\bar\psi + \gamma_{01} - \pi)$ at $r = \bar r$.

As before, we take rigid motions of $U$ and $V$ to match the interfaces at $r = -\bar r$ using the Lagrange multiplier $\lambda$ to obtain the physical surfaces denoted by $u$ and $v$.  Then when evaluated at $r = \bar r$, $\bar v$ in general is not equal to $\bar w$ in this case either, so a similar $F(\bar\psi)$ is constructed here, and a zero is found numerically using Matlab's fzero.

As in the earlier case, with this critical value of $\bar\psi$, we have matched $u$, $v$, and $w$ at a prescribed $\bar r> 0$.  As before, in general, the enclosed volume of the drop just formed will not match the prescribed volume of the drop.

To compute the enclosed volume for the drop bound to a wall, following the above computation, we obtain
\begin{equation}
\mbox{Vol}_V = \pi\bar V \left( R^2 - {\bar r}^2 \right) - \frac{2\pi}{\kappa_{12}} \left( R\sin\left( \frac{\pi}{2} - \gamma^2_{1p} \right)  - \bar r \sin(-\bar\psi) \right),
\end{equation}
where we use  $-\bar\psi$ at $\bar r$ as the angle there is derived from the left side of the drop, though the volume is more naturally derived from the right side.  We also have 
\begin{equation}
\mbox{Vol}_U = \pi\bar U\left( R^2 - {\bar r}^2 \right) - \frac{2\pi}{\kappa_{01}}\left( R\sin\left( \gamma^1_{0p} - \frac{\pi}{2} \right)  - \bar r \sin\left( \bar\psi - \gamma_{02} \right) \right).
\end{equation}
also giving the form $\mbox{Vol} = \mbox{Vol}_U + \mbox{Vol}_V$ in this case.

\begin{figure}[t]
	\centering
	\includegraphics[scale=.2]{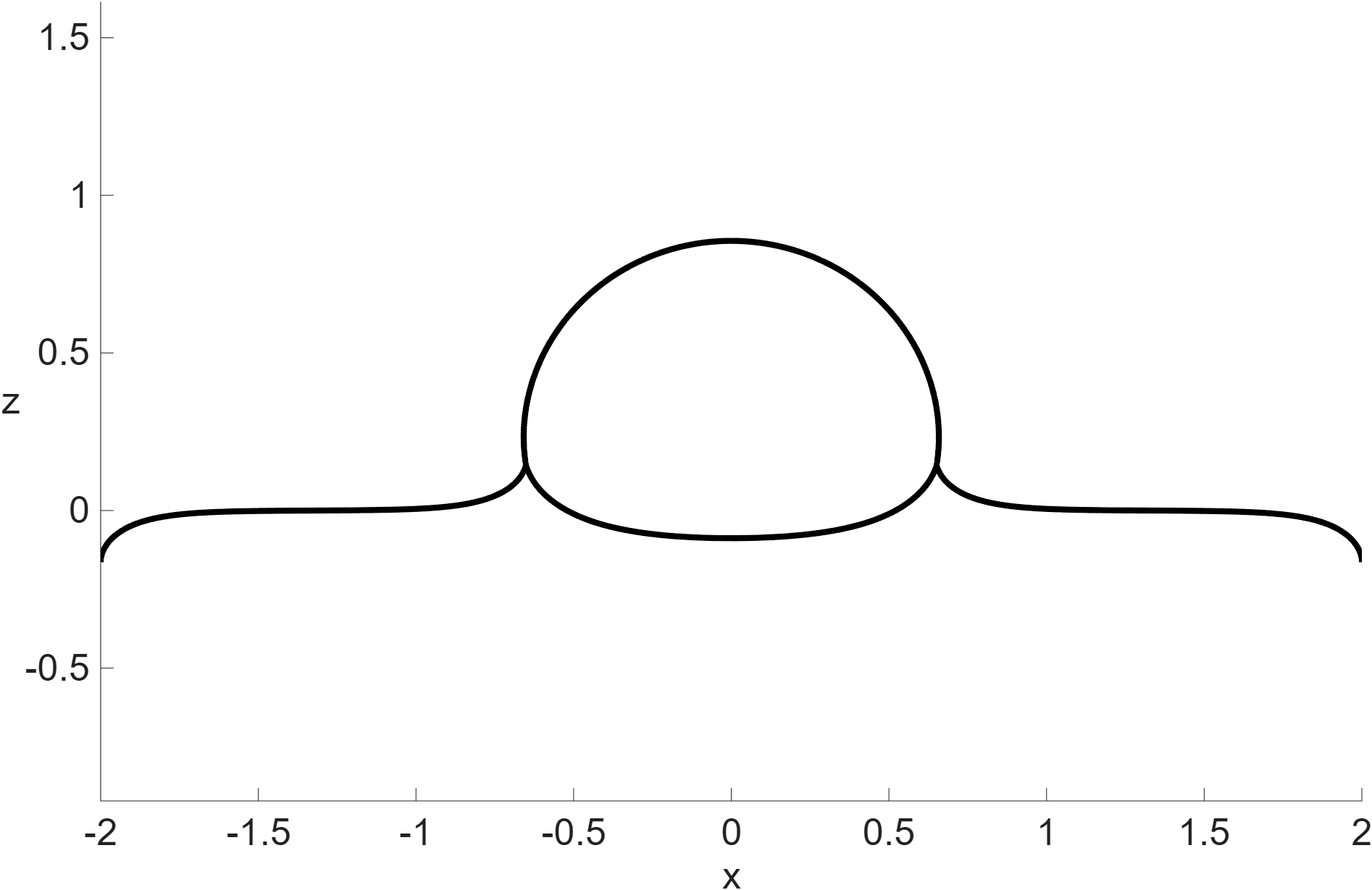}
	\qquad
	\includegraphics[scale=.2]{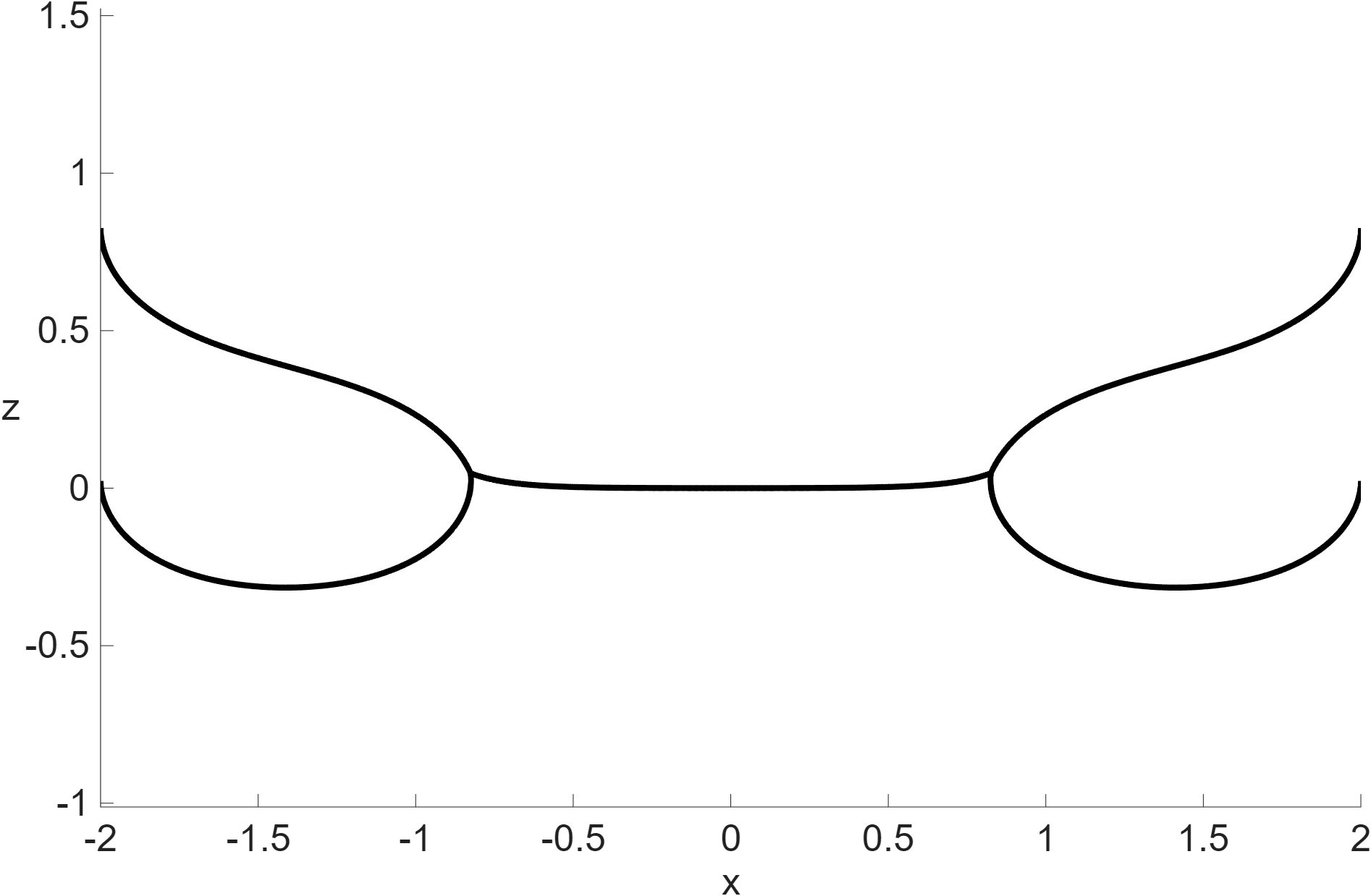}
	\caption{Displayed are analogues of the symmetric floating drops in $\mathbb{R}^2$.  On the left is a centrally located drop with the physical parameters that determine this drop being an area of 1, tube radius of $X=2$, densities $\rho_1 = 0.4$ and  $\rho_2 = 15$, surface tensions $\sigma_{01} = 7.9, \sigma_{02} = 2$, and $\sigma_{12} = 6.1$, and plate angle $\gamma^2_{0p} = \pi - 0.01$.  On the right is a wall-bound drop with the physical parameters that determine this drop being an area of 1.5 split evenly on each wall, tube radius of $X=2$, densities $\rho_1 = 7.5$ and  $\rho_2 = 15$, surface tensions $\sigma_{01} = 7.5, \sigma_{02} = 2.5$, and $\sigma_{12} = 6$, and plate angles $\gamma^1_{0p} = 0.01$ and $\gamma^2_{1p} = 0.01$.}
	\label{fig:2dSymmetricDrops}
\end{figure}

\begin{figure}[h!]
	\centering
	\includegraphics[scale=.2]{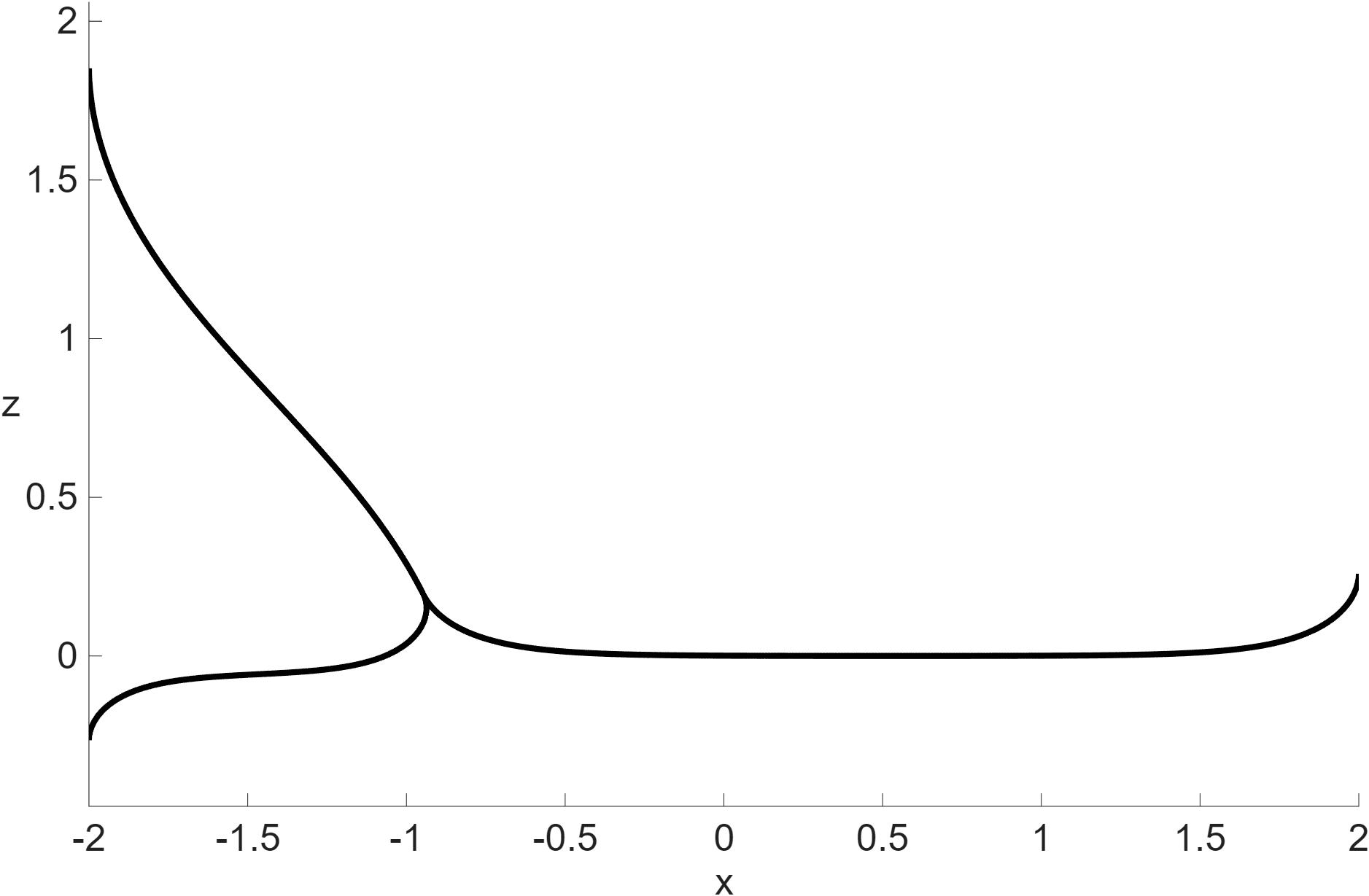}
	\qquad
	\includegraphics[scale=.2]{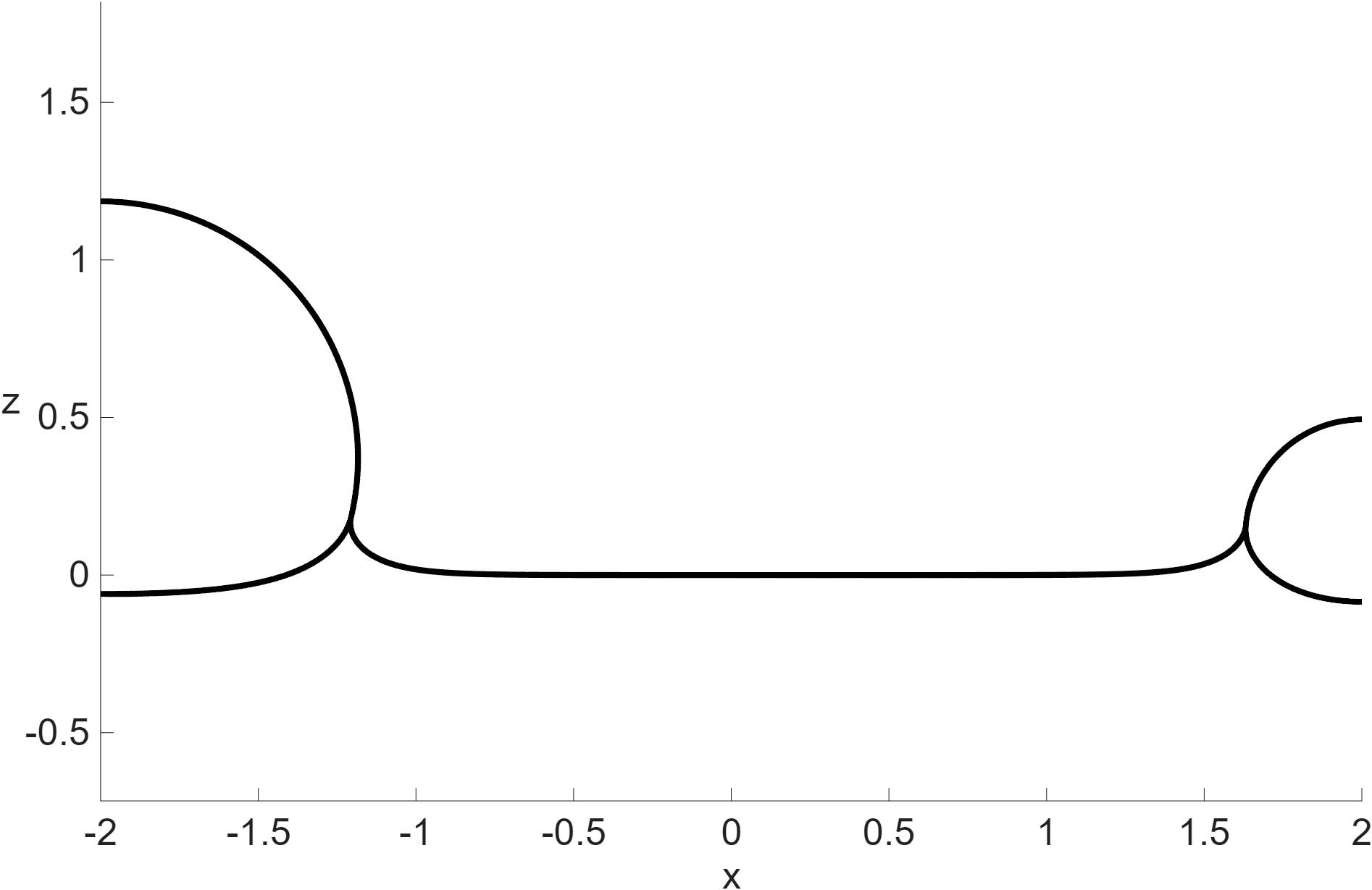}
	\caption{Asymmetric floating drops in $\mathbb{R}^2$.  On the left is a drop adjacent to the left wall with the physical parameters that determine this drop being an area of 1, tube radius of $X=2$, densities $\rho_1 = 1, \rho_2 = 15$, surface tensions $\sigma_{01} = 7.9999, \sigma_{02} = 5$, and $\sigma_{12} = 3.0001$, and plate angles $\gamma^1_{0p} = 0.01$ and $\gamma^2_{1p} = \pi - 0.01$.  On the right is a drop split with 1/6 of the  area of liquid adjacent to the right wall, and the remaining area of fluid adjacent to the left wall. The physical parameters are a total area of 1, tube radius of $X=2$, densities $\rho_1 = 0.01$ and $\rho_2 = 15$, surface tensions $\sigma_{01} = 7.9999, \sigma_{02} = 2.0001$, and $\sigma_{12} = 6$, and plate angles $\gamma^2_{0p} = \pi/2$ and $\gamma^2_{1p} = \pi/2$.}
	\label{fig:2dAsymmetricDrops}
\end{figure}

Then we again have a second fzero module where we match the computed volume to the prescribed volume by varying $\bar r$.  The output of these two nested loops gives a solution to the floating drop problem with a wall-bound drop.

\subsection{Floating drops problems in $\mathbb{R}^2$}
\label{subsec:2d}

The above processes can also be carried out for $\mathbb{R}^2$.  We will have more cases to consider here.  Of course, we have the centrally located drop and the analogue of the symmetric wall-bound drop as above which is actually two drops of equal area, each of which is bound to an opposing wall, and these are shown in Figure~\ref{fig:2dSymmetricDrops}.  We are not restricted to symmetric interfaces here, as asymmetric configurations are possible to compute.  One such configuration being a wall-bound to a single wall, and we can also have split drops with different areas of the drop attached to each wall, and these are shown in Figure~\ref{fig:2dAsymmetricDrops}.  

There may be other configurations possible, but those are the ones we consider in this work.  The notation changes so that  the boundary of the container at $x = -X$ and $x = X$, and the free boundary is located at $x = \pm\bar x$ for the centrally located drop, or some $x = \bar x$ generically otherwise. 
The main changes are the area formulas, where the components become 
\begin{eqnarray}
	\mbox{Area}_V &=& 2 \int^{\bar x}_0 \left( \bar V - V \right)\, dx \nonumber\\
	&=& 2\bar x \bar V - \frac{2}{\kappa_{12}}\int^\ell_0 \frac{d\psi}{ds}\frac{dx}{ds}\, ds \nonumber\\
	&=& 2\bar x \bar V - \frac{2}{\kappa_{12}}\sin\bar \psi,
\end{eqnarray}
\begin{equation}
	\mbox{Area}_U = 2\bar x \bar U - \frac{2}{\kappa_{01}}\sin\left(\gamma_{02} - \bar \psi \right),
\end{equation}
for the central drop.  For the wall-bound drop on the right we have 
\begin{eqnarray}
	\mbox{Area}_V &=& \int^X_{\bar x} \left( \bar V - V \right)\, dx \nonumber\\
	&=& (X - \bar x) \bar V - \frac{1}{\kappa_{12}}\left( \sin\left( \frac{\pi}{2} - \gamma^2_{1p} \right) - \sin\left( - \bar \psi \right) \right),
\end{eqnarray}
and
\begin{equation}
	\mbox{Area}_U = (X - \bar x) \bar U - \frac{1}{\kappa_{01}}\left( \sin\left( \gamma^1_{0p} - \frac{\pi}{2} \right) - \sin\left(  \bar\psi - \gamma_{02} \right) \right),
\end{equation}
and for a wall-bound drop on the left where the interfaces span from $x = -X$ to  the free boundary at $x = \bar x$, we have
\begin{equation}
	\mbox{Area}_V =  (\bar x + X) \bar V - \frac{1}{\kappa_{12}}\left( \sin\bar \psi - \sin \left( \gamma^2_{1p} - \frac{\pi}{2} \right)  \right),
\end{equation}
and
\begin{equation}
	\mbox{Area}_U =  (\bar x + X) \bar U - \frac{1}{\kappa_{01}}\left( \sin\left( \gamma_{02} - \bar \psi \right)  - \sin \left( \frac{\pi}{2} - \gamma^1_{0p} \right)  \right).
\end{equation}
Of course, if there is a split drop that has an area attached to each wall, we can use these formulas to compute those areas.  This split drop has two different free boundary triple junction points, and they are varied independently to achieve the prescribed area on each side of the container.

\subsection{Parameter discussion}

In this section we have discussed the model of a floating drop, and there are a number of parameters that match the model to a particular physical configuration.  The list of the nine available parameters is $\mbox{Vol}$ ($\mbox{Area}$), the volume (area) of the drop, the container size indicated by $R$ or $X$, the two densities $\rho_1$ and $\rho_2$, the three surface tensions $\sigma_{01},\sigma_{02},$ and $\sigma_{12}$, and the two free wetting energies $-\sigma_{01}\cos\gamma^1_{0p}$ and $-\sigma_{02}\cos\gamma^2_{0p}$.  Since our model only relies on the difference in the densities, we are able to assume $\rho_0 = 0$, and we also use the dependence of the third wetting energy $\sigma_{12}\cos\gamma^2_{1p}$ on the two free wetting energy  parameters.   Our treatment of these parameters is decidedly abstract, and we make no attempt to fit our solutions to experimental data.  We do explore this nine-dimensional parameter space in what follows.  In the next section, we start with an example that is potentially experimentally reproducible by increasing the volume of the drop.  It would be interesting if an experimentalist could reproduce our results physically.  In the remainder of the paper we will focus on other parameter studies that we found to give interesting results.  In our time working with these problems, we have found that there are four groupings of parameters.  The bulk parameters are the container size and the drop volume (area), and we have chosen to fix $R=2$ and $X=2$ while varying the drop volume (area).  The two terms related to the gravitational energy are seen in the two densities, and we fix $\rho_2 = 15$ and consider a variety of choices of $\rho_1$.  The contact angles at the triple junction free boundary (boundaries) are derived by the three surface tensions, and we arbitrarily normalize these to add up to a value of 16.  We often choose values of surface tensions to get $\gamma_{02}$ near $\pi$, as we have found such edge cases make more striking examples.  Finally, the wetting energies do depend on the surface tensions, but \eqref{eqn:plate} allows us to fix two plate contact angles to derive the third.  This relation is always present even when the drop does not touch the wall, and the physical wetting energy constants are present even if they are not part of the energy for a particular configuration.  While we report both typical and interesting cases of the explorations we have made into this somewhat large parameter space, there are complicated interactions between the parameters, and we make no claim that we have exhausted all of the interesting cases. 

\section{Floating drops in $\mathbb{R}^3$}
\label{sec:3D}

Here we discuss the remaining details needed to formulate the potential energy for configurations in $\mathbb{R}^3$.  We also explore an example that illustrates a process that could lead to a physical experiment for verification of our results, and then we give some examples from  our basic groups of physical parameters as well as an interesting geometrical symmetry in the parameter space.

\begin{figure}[h]
	\centering
	\includegraphics[scale=.2]{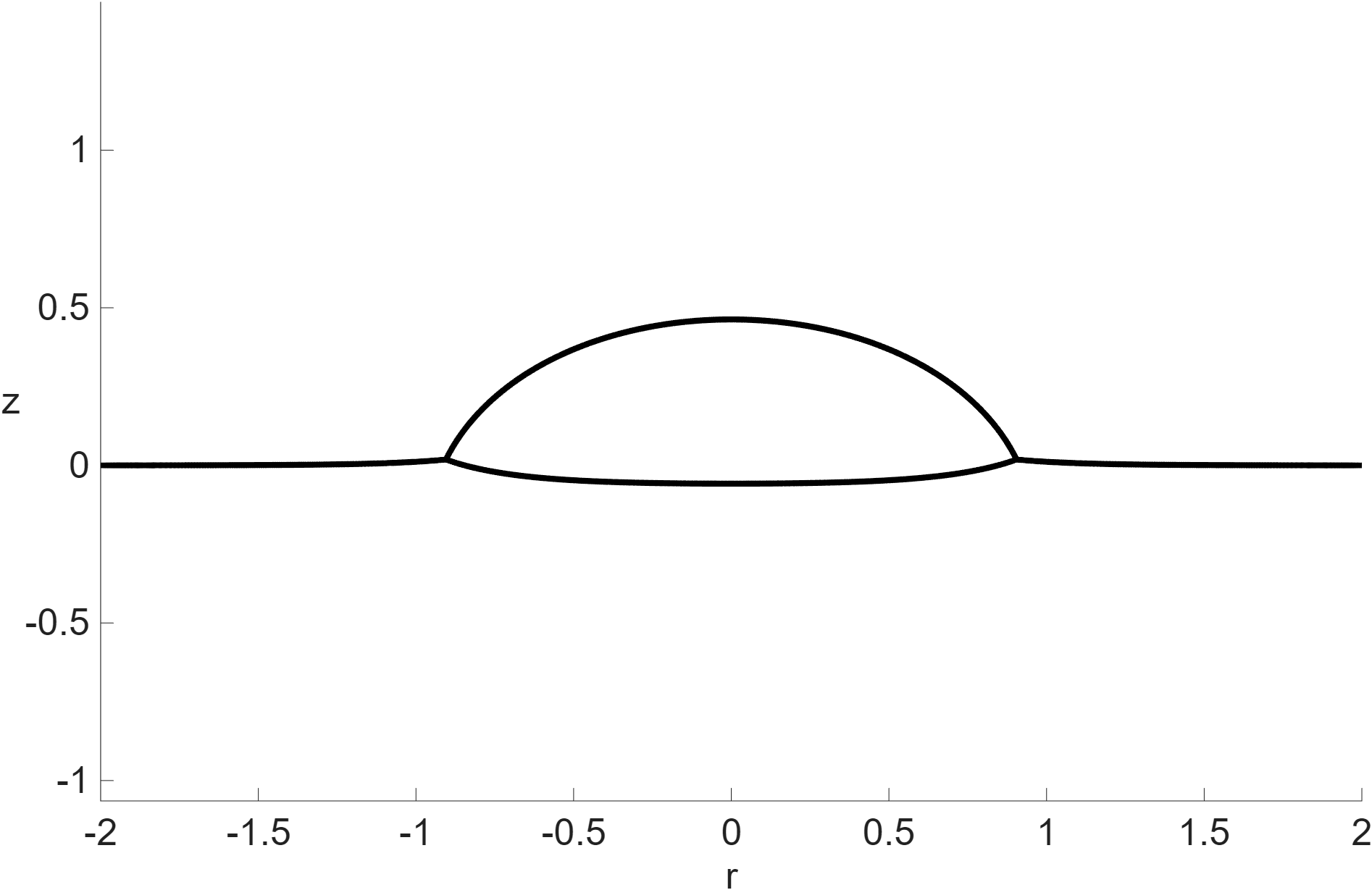}
	\qquad
	\includegraphics[scale=.2]{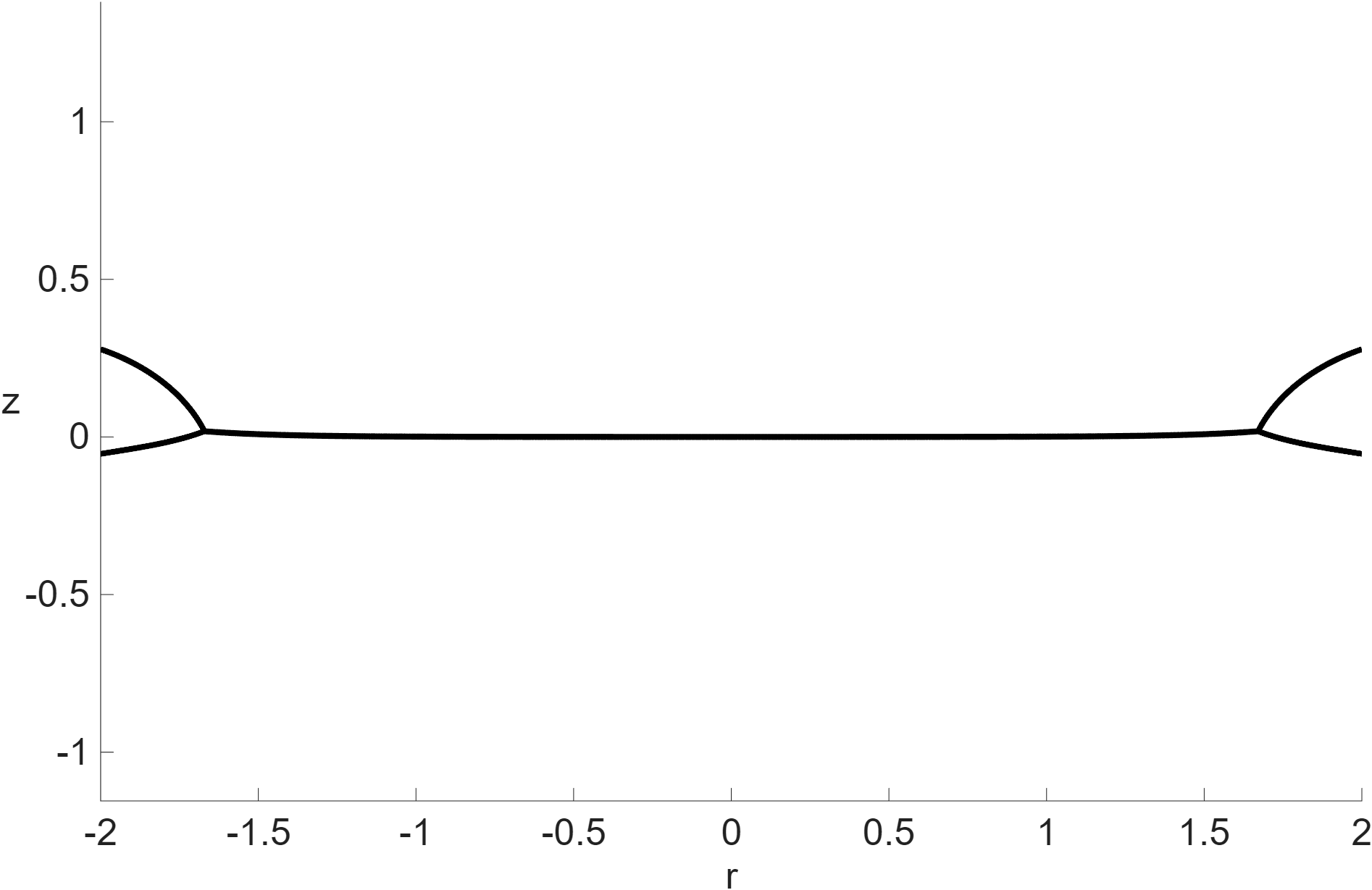}
	\caption{A pair of axially symmetric floating drops with the same parameters and a volume of 0.8.  The potential energy of the wall-bound drop is lower than that of the centrally located drop.  The physical parameters that determine these drops are a  tube radius of $R=2$, densities $\rho_1 = 1$ and $\rho_2 = 15$, surface tensions $\sigma_{01} = 3, \sigma_{02} = 7$, and $\sigma_{12} = 6$, and plate angles 
		$\gamma^2_{0p} = \pi/2$, and $\gamma^2_{1p} = 1.73324$.}
	\label{fig:3dCentralAndWallDropSmall}
\end{figure}

\begin{figure}[h]
	\centering
	\includegraphics[scale=.2]{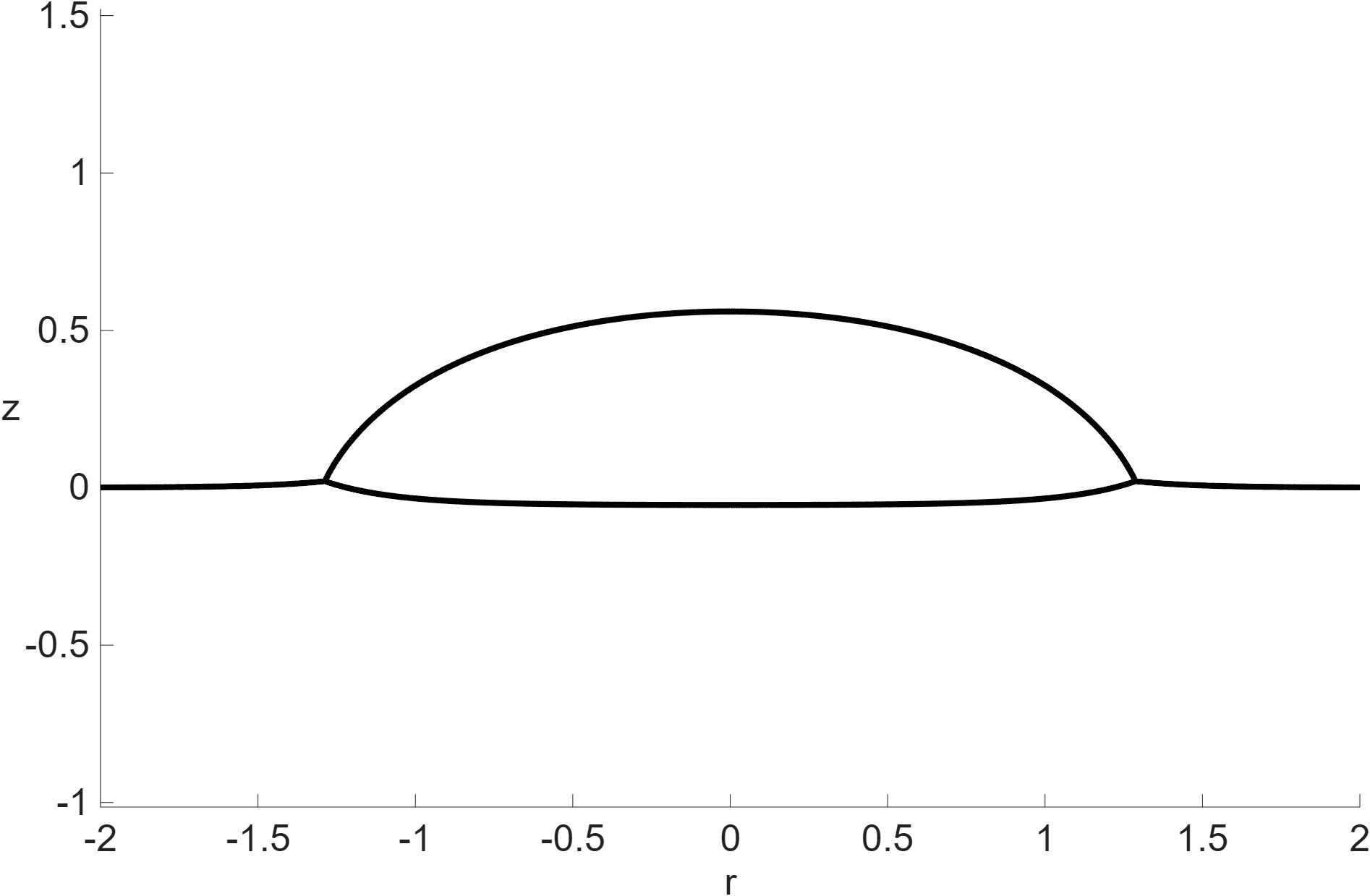}
	\qquad
	\includegraphics[scale=.2]{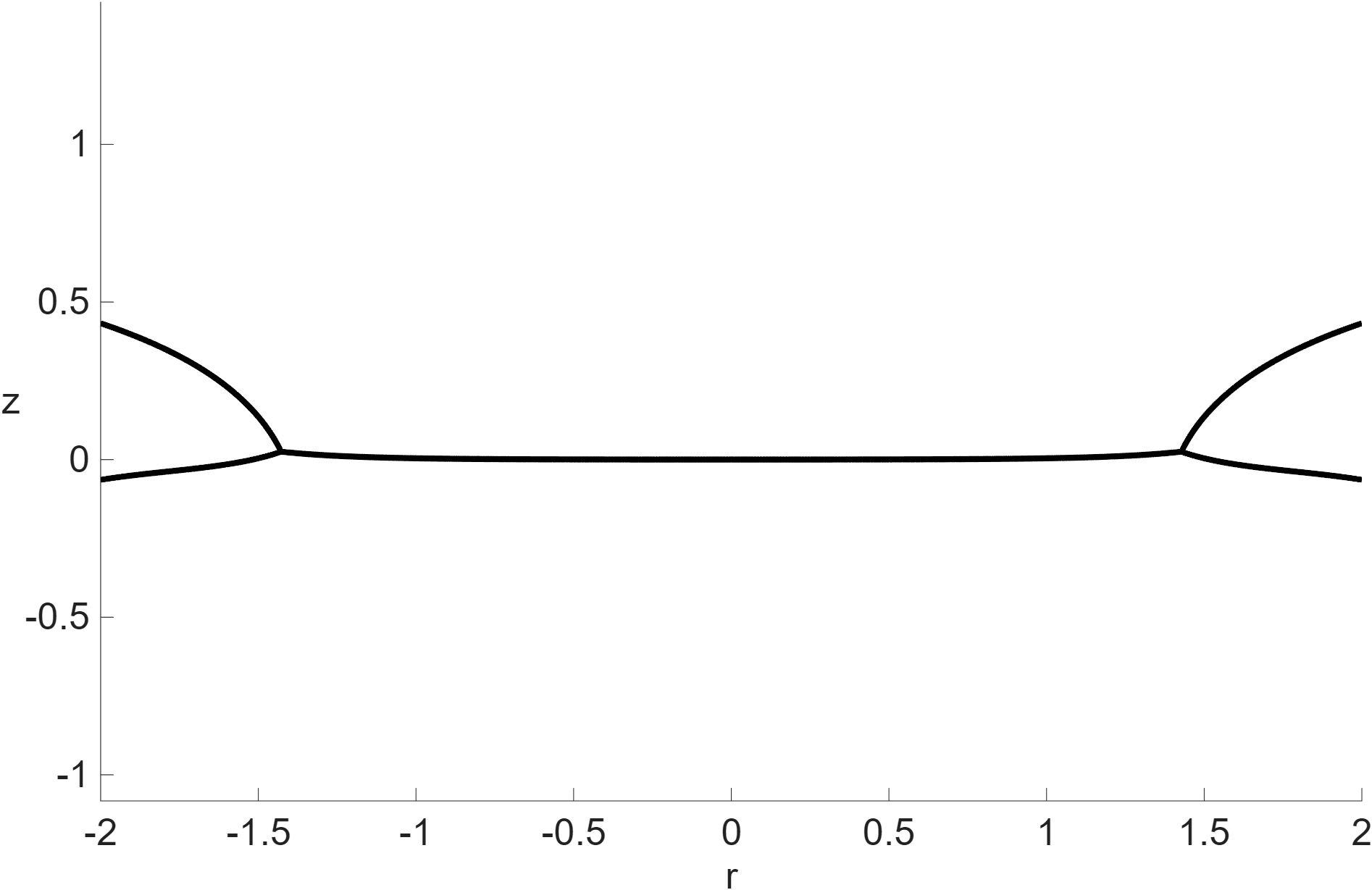}
	\caption{A similar pair of  axially symmetric  floating drops with the same parameters as those in Figure~\ref{fig:3dCentralAndWallDropSmall}, except the drop volume is increased to 2.  The potential energy of the centrally located drop is lower than that of the wall-bound drop.}
	\label{fig:3dCentralAndWallDropBig}
\end{figure}
\subsection{An example suitable for physical verification}
\label{sec:physical}

In Figure~\ref{fig:3dCentralAndWallDropSmall} there are two different floating drop configurations that have the same parameters and a prescribed volume of 0.8.  One drop is centrally located and the other is bound to the wall of the cylindrical container, and both are radially symmetric, giving an example of the non-uniqueness of the floating drop problem.  In every choice of elements in our parameter space that we tested we have observed this non-uniqueness.  Since both configurations are solutions to the Euler Lagrange equations as detailed above, the natural question is to determine the configuration with lower energy.  

We need some details of how to compute the energy, which we will turn to shortly.  First, we observe that the resulting energy of the centrally located drop is 95.322 and the energy of the wall-bound drop is 95.2593, so the wall-bound drop achieves a lower potential energy.  Then in Figure~\ref{fig:3dCentralAndWallDropBig} the situation is reversed, while the parameters remain the same except the drop volume is increased to 2, and the energy of the centrally located drop is 101.9708 however the energy of the wall-bound drop is 102.8821 and the centrally located drop achieves a lower potential energy.

\begin{figure}[t]
	\centering
	\includegraphics[scale=.2]{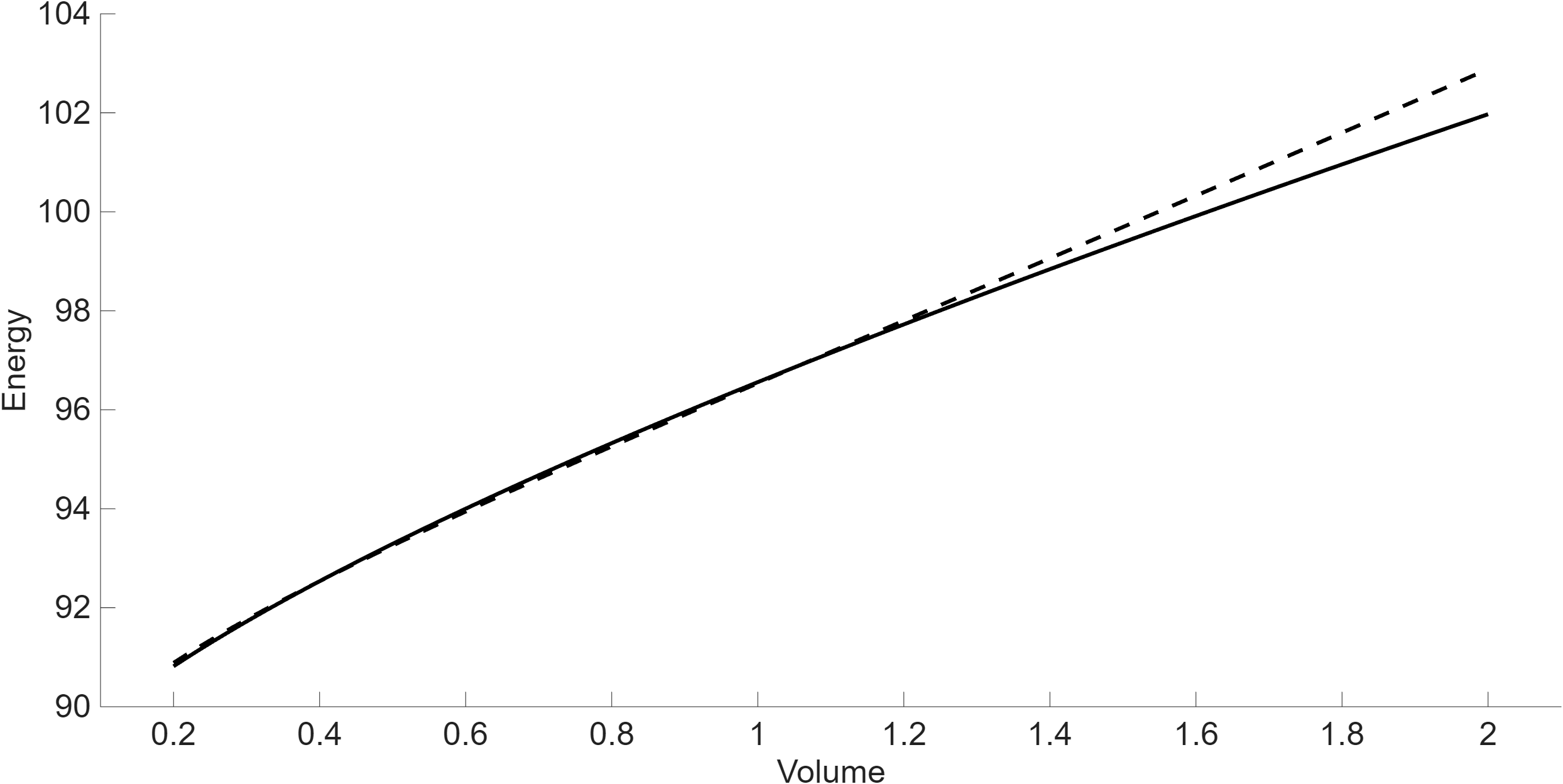}
	\caption{Comparing the energies of the centrally located drop (solid) and the wall-bound drop (dashed) as the volume of the drop is continuously increased.}
	\label{fig:3dVolumeEnergy}
\end{figure}

In computing the potential energy for these configurations, there are terms for the surface area, gravitational potential, and wetting.  To compute the surface areas of the component interfaces in $\mathbb{R}^3$, we observe that the standard formula gives
\begin{equation}
S = 2\pi \int^b_a r\, ds
\end{equation}
for solutions to the ODEs for some $a$ and $b$ when $r > 0$.  If the interface does not intersect the vertical axis, then we simply set $a = 0$ and $b = 2\ell$ which is the total arc-length of our generating curve.  Since we have computed the solution $r$ using Chebyshev points, it is natural to use a Clenshaw-Curtis quadrature to compute these integrals.    Chebfun \cite{Chebfun} is used to build the weights for these nodes.  See Trefethen \cite{ATAP} for a discussion of the error bounds on these approximations.  We will say that in our case there are $n + 1 = 14$ Chebyshev points in a typical surface, and the error of an $(n+1)$-point Clenshaw-Curtis quadrature $I_n$ for the approximation of the integral $I = \int f(z) \, dz$ satisfies
\begin{equation}
|I - I_n| \leq \frac{64}{15}\frac{M \rho^{1 - n}}{\rho^2 - 1}
\end{equation}  
for some $M$ with the integrand $|f(z)| \leq M$, and $\rho$ from the associated Bernstein ellipse $E_\rho$.
If our surface spans the axis of symmetry, then we compute 
\begin{equation}
S = \pi \int^\ell_{-\ell}  |r|\, ds
\end{equation}
using a halved double cover and the weights applied across the whole section, which is where we have our Chebyshev points.  

Similarly, the gravitational potential energy of an interface $w$ that spans the vertical axis is given by
\begin{equation}
G = 2\pi\int^{\bar r}_0 \rho g\sgn(w)r w^2\, dr = 2\pi \int^\ell_0 \rho g\sgn(w)r w^2 \cos\psi\, ds,
\end{equation}
for an appropriate density $\rho$, and with the usual $\mbox{sgn}(w)$ function indicating the sign of $w$ so that the gravitational potential is measured with respect to the reference level.  This formula is derived here for the case when $w$ is a graph over $B$, however, the arc-length form extends to parametric configurations that are not graphs over $B$.  

These formulas are given generically, and are readily adapted to each component interface in our centrally located and wall-bound configurations.

The wetting energy for any surface is taken to be above or below the reference height $z\equiv0$.  If a fluid wets the wall, the wetting energy is simply the interface height on the boundary multiplied by $2\pi R$ and the relevant weight for that fluid.

With these details established, and motivated by the examples in Figures~\ref{fig:3dCentralAndWallDropSmall} and \ref{fig:3dCentralAndWallDropBig}, we can now describe an experiment that could be physically conducted.  We start with a small drop volume that is centrally located and continuously inject more volume into the drop, and while we increase the volume we measure the energy of the configuration.  Then we repeat this process for the symmetric wall-bound drop.  The results are collected in Figure~\ref{fig:3dVolumeEnergy}.  This is the first of a number of experiments of this type.  We are continuously changing the volumes of both the centrally located drop and the wall-bound drop and measuring the energy of each.  Later we will consider other parameters besides volume.  A careful view of this figure shows that there are two volumes where the centrally located drop has the same potential energy as the wall-bound drop.  This is the first time there has been observed non-uniqueness of (presumed) energy minimizers in the floating drop problem.  Of course, there are also possibly non-symmetric solutions to the floating drop problem, and we have not produced those examples here.

\begin{figure}[t]
	\centering
	\includegraphics[scale=.2]{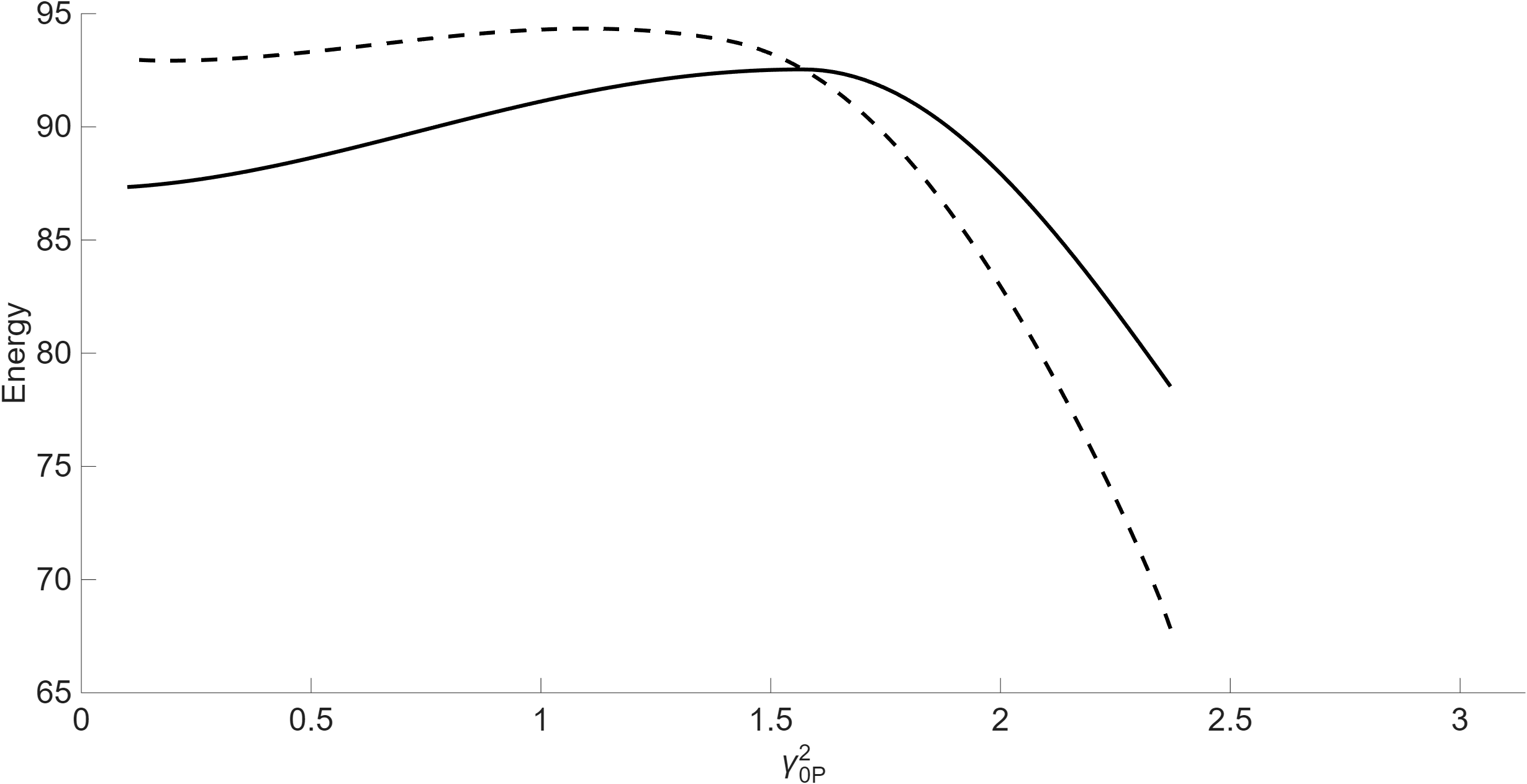}
	\caption{Comparing the energies of the centrally located drop (solid) and the wall-bound drop (dashed) in $\mathbb{R}^3$.  We are varying the contact angle $\gamma^2_{0p}$, which roughly determines if $S_{02}$ is concave up.  The physical parameters are a volume of $0.4$ with tube radius $R = 2$, densities $\rho_1 = 1$ and  $\rho_2 = 15$, surface tensions $\sigma_{01} = 3, \sigma_{02} = 7$, and $\sigma_{12} = 6$, plate angle $\gamma^1_{0p} = 1.24139754239$.}
	\label{fig:3dCaseCPlate}
\end{figure}

\subsection{More examples}
\label{sec:3dmore}

Next, we turn to the question of the balance between the gravitational potential energy and wetting energies.  A guiding heuristic is that if the capillary tube configuration without a drop added gives an interface that is concave up, which means $0 < \gamma^2_{0p} < \pi/2$, then adding a drop will lead to a centrally located drop as the energy minimizer.  The physical reasoning is that the drop will descend to the bottom of the interface $S_{02}$ in order to reduce the gravitational potential.  The converse of this is when the interface is concave down and $\pi/2 < \gamma^2_{0p} < \pi$ and then the heuristic is that the wall-bound drop would be the energy minimizer.  We see these ideas in Figure~\ref{fig:3dCaseCPlate}, where the centrally located drop has lower energy for all $\gamma^2_{0p}$ smaller than $\pi/2$, and when $\gamma^2_{0p}>\pi/2$ the wall-bound drop has the smaller value of energy.    We will pick this topic up again in Section~\ref{sec:2d}.

\begin{figure}[t]
	\centering
	\includegraphics[scale=.2]{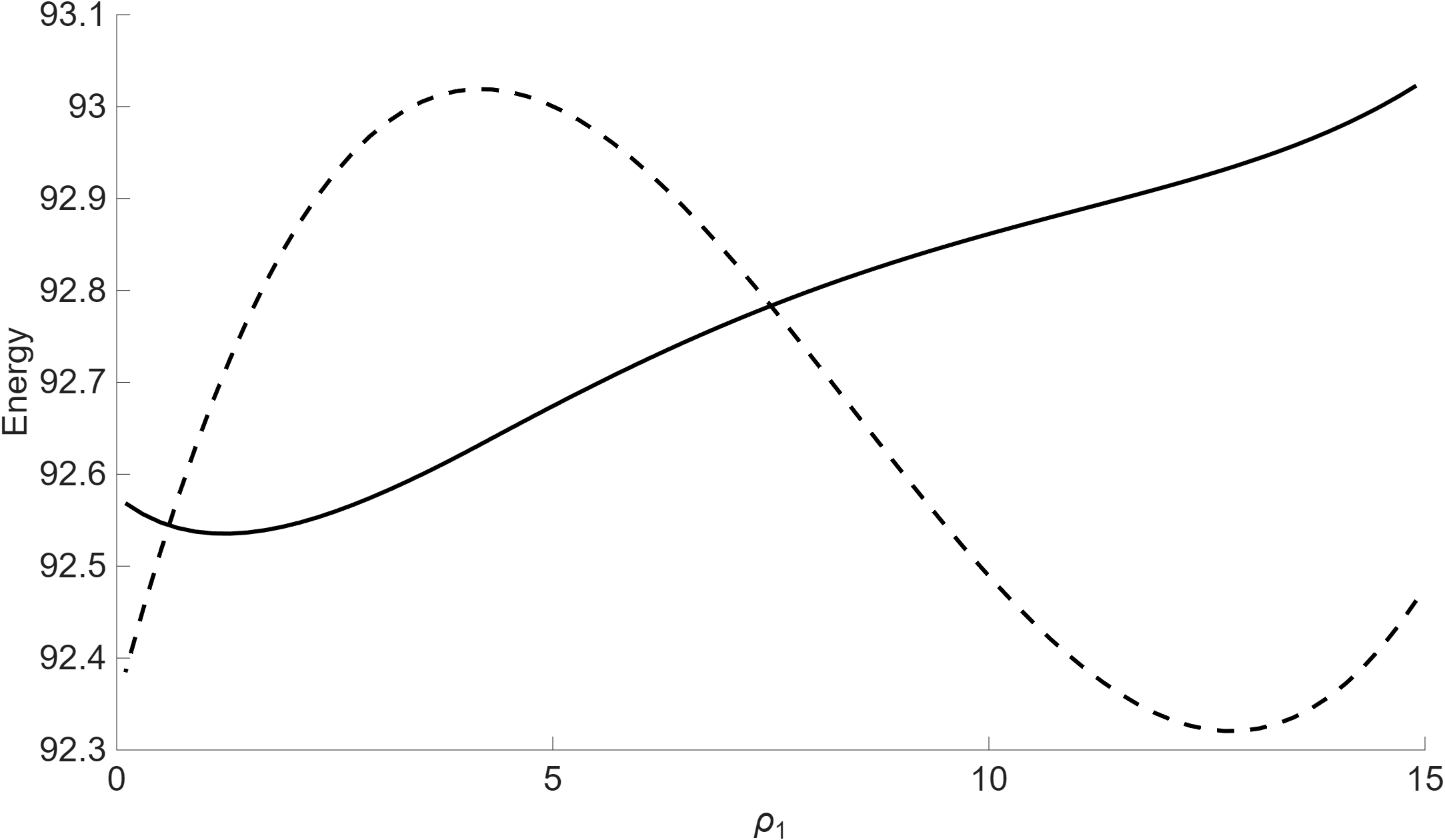}
	\caption{Comparing the energies of the centrally located drop (solid) and the wall-bound drop (dashed).  We are varying the density $\rho_1$ in this experiment.  The physical parameters are a volume of $0.4$ with tube radius $R = 2$, density $\rho_2 = 15$, surface tensions $\sigma_{01} = 3, \sigma_{02} = 7$, and $\sigma_{12} = 6$, plate angles $\gamma^2_{0p} = \pi/2$ and $\gamma^2_{1p} = 1.727954$.}
	\label{fig:3dCaseCDensity}
\end{figure}

In Figure~\ref{fig:3dCaseCDensity} we show an experiment where the density of the drop is increased from $\rho_1 = 0$ to $\rho_1 = \rho_2 = 15$.  Here there are two values of $\rho_1$ where the energy values are the same, and this is quite common in our parameter studies in $\mathbb{R}^3$.  In this experiment and many that follow, we focus on the neutrally wetting case where $\gamma^2_{0p} = \pi/2$ and there is no driving concavity of the fluid in the tube without a drop present.

The fourth type of groupings of the energy terms would be the surface tensions themselves.  We save an experiment involving $\gamma_{02}$ for Section~\ref{sec:2d} as our explorations have not turned up a difference in the dimension of the problem for experiments of this type.  

\begin{figure}[t]
	\centering
	\includegraphics[scale=.2]{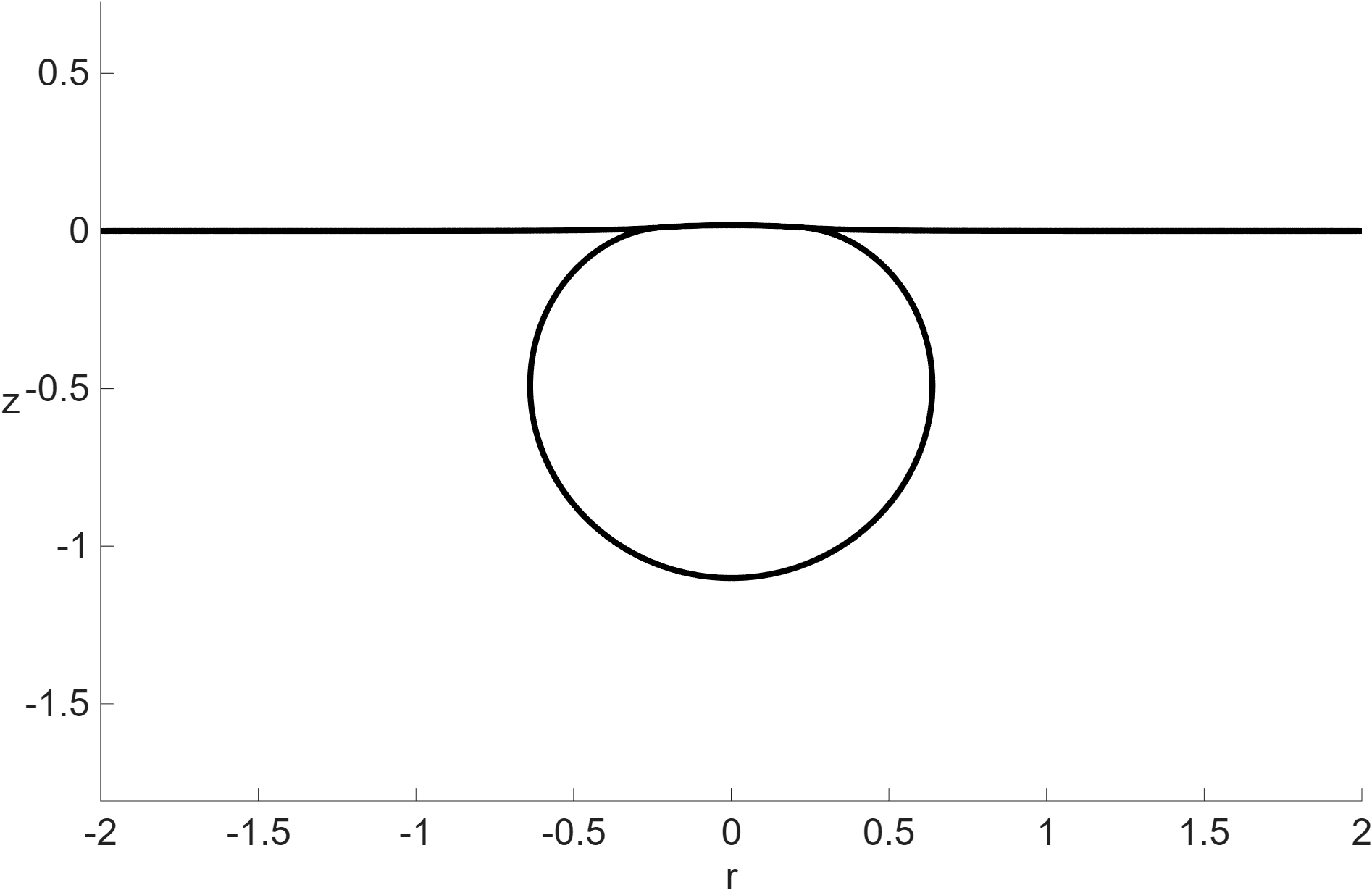}
	\qquad
	\includegraphics[scale=.2]{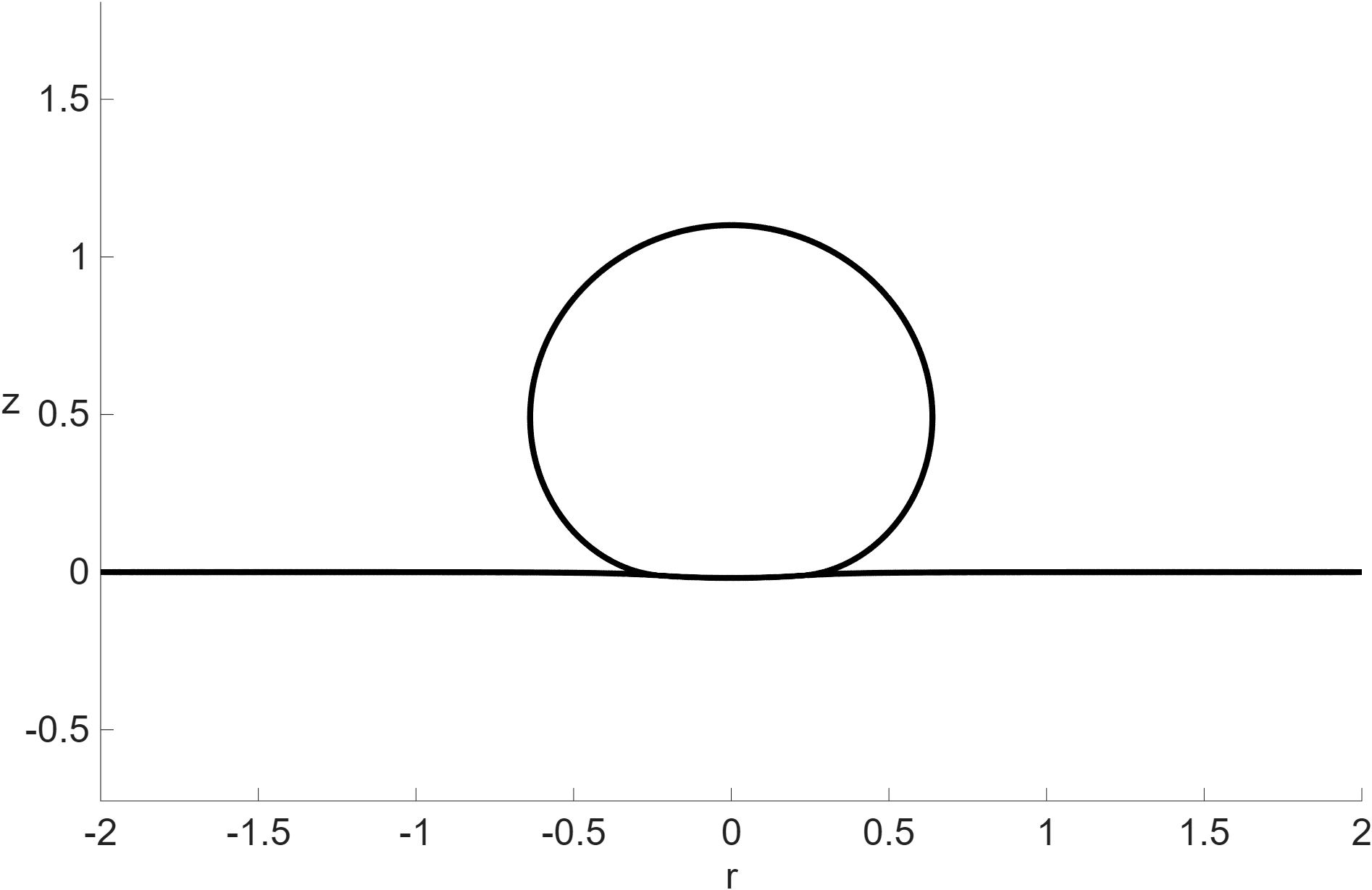}
	\caption{Here are two drops displaying a reflection symmetry of the parameters.  For both drops, the volume is 1, tube radius of $R=2$, the density of $E_2$ is $\rho_2 = 15$, a common surface tension of $\sigma_{02} = 6$, and plate angle $\gamma^2_{0p} = \pi/2$.  On the left $\rho_1 = 14.9$, $\sigma_{01} = 7.9999$, $\sigma_{12} = 2.0001$.  On the right $\rho_1 = 0.1$, $\sigma_{01} = 2.0001$, $\sigma_{12} = 7.9999$.}
	\label{fig:3dSymmetricParameters}
\end{figure}

We do, however, finish our discussion of drops in $\mathbb{R}^3$ with a general observation on the symmetry of the geometry of the drops with respect to a certain symmetry in the parameters. Given fixed $\gamma^2_{0p} = \pi/2$ and any admissible $\sigma_{02}$, as well as any admissible volumes, tube widths, densities $\rho_2$, then if the surface tensions $\sigma_{01} = a$ and $\sigma_{12} = b$ are also admissible, and the density $\rho_1 = d \in[0,\rho_2]$, then the solution is vertically reflected when choosing the surface tensions $\sigma_{01} = b$ and $\sigma_{12} = a$ and density $\rho_1 = \rho_2 - d$.  We show an example of this in Figure~\ref{fig:3dSymmetricParameters}.

\section{Floating drops in $\mathbb{R}^2$}
\label{sec:2d}

Here we discuss the remaining details needed to formulate the potential energy for configurations in $\mathbb{R}^2$.  In the following subsections, among other results, we give a parameter study we omitted when we considered floating drops in $\mathbb{R}^3$, we show how the heuristic guide of the concavity driving whether the centrally located drop is the energy minimizer is false in general, and we conclude with examples that highlight symmetry breaking.

In Subsection~\ref{subsec:2d} the details of computing floating drops in $\mathbb{R}^2$ were finished.  In order to compute the energy of these configurations, we need three quantities: surface area becomes interface length, wetted area becomes wetted length (if any), and gravitational potential.  The interface length here is  merely the total length of the generating curve, given by $2\ell$, and output by the solver as a part of obtaining the interface.  Then multiplying that quantity by the appropriate surface tension gives the potential energy due to surface tension for that interface.  As before, the wetting energy of any liquid wetting a wall is obtained by using the height of the interface there.  In this case, the height above the reference height $z\equiv0$ there is simply multiplied by an appropriate physical constant.  For the gravitational potential, a generic formula is given by
\begin{equation}
	G = \int^b_a \rho g\sgn(w) w^2\, dr = \int_{-\ell}^\ell \rho g\sgn(w) w^2\cos\psi\, ds
\end{equation}
for some generic $a < b$ and an appropriate density $\rho$.  As before, the form on the left assumes the interface is a graph over $B$, but the form on the right is given by a parametrization by arc-length, and holds in the parametric form despite the details being omitted here.  These quantities are computed from the Chebyshev points along the interface obtained from the solver, and as before, the Clenshaw-Curtis quadrature weights are employed to compute the integrals.  With these details in hand, we are able to compute the potential energy for the configurations we consider.

\begin{figure}[t]
	\centering
	\includegraphics[scale=.2]{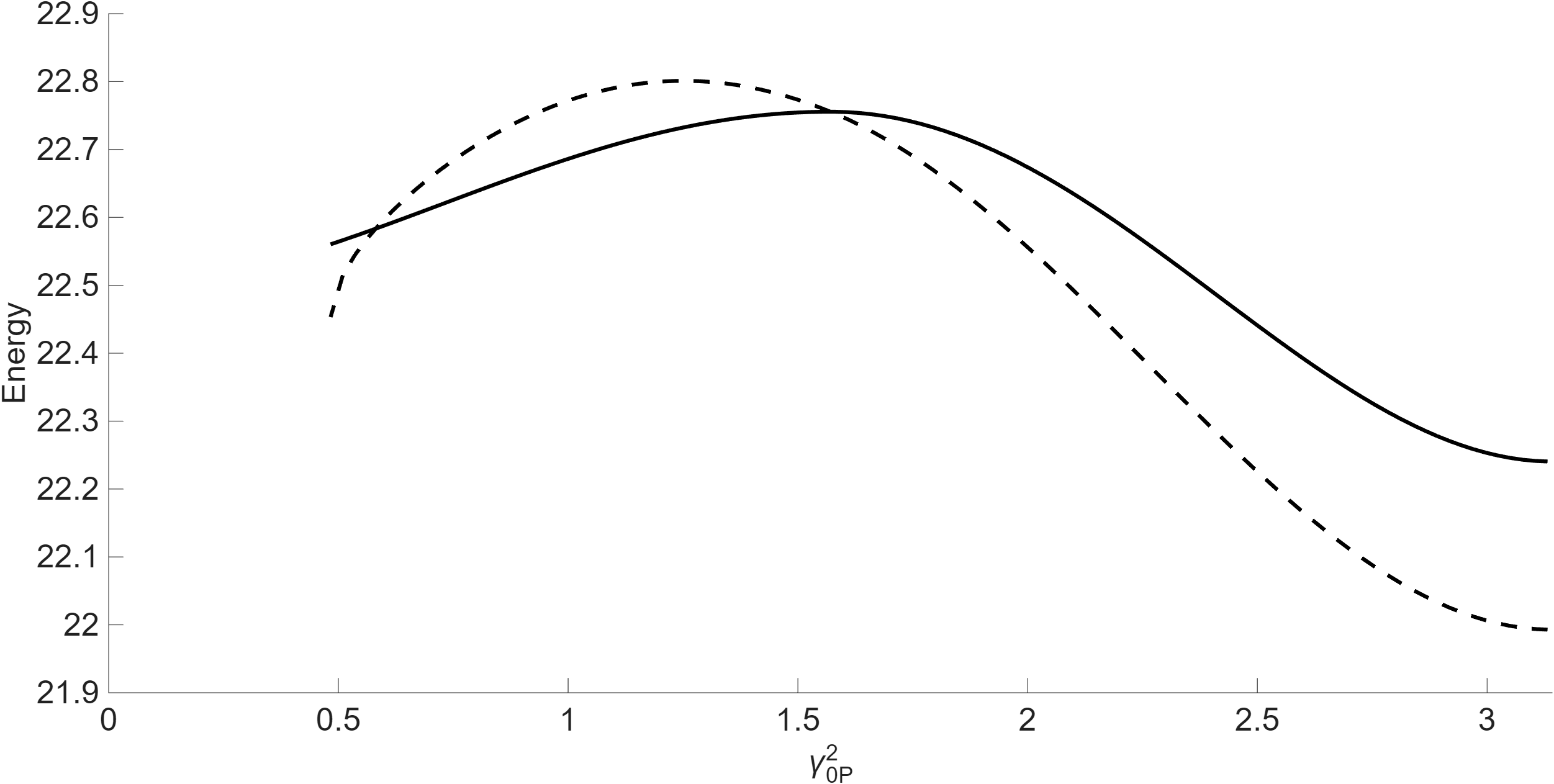}
	\caption{Here is a range of plate angles $\gamma^2_{0p}$ where the heuristic breaks.  On the left there is an atypical second crossing of the curves away from $\pi/2$.  
		The physical parameters that determine these are an area  of $0.4$, tube radius of $X = 2$, densities $\rho_1 = 5$ and $\rho_2 = 15$, surface tensions $\sigma_{01} = 7.9999, \sigma_{02} = 2.0001$, and $\sigma_{12} = 6$. One plate angle is fixed as $\gamma^1_{0p} = 2.1277$, and the third is determined by the governing equation.}
	\label{fig:2dantiheuristic}
\end{figure}

\begin{figure}[t]
	\centering
	\includegraphics[scale=.2]{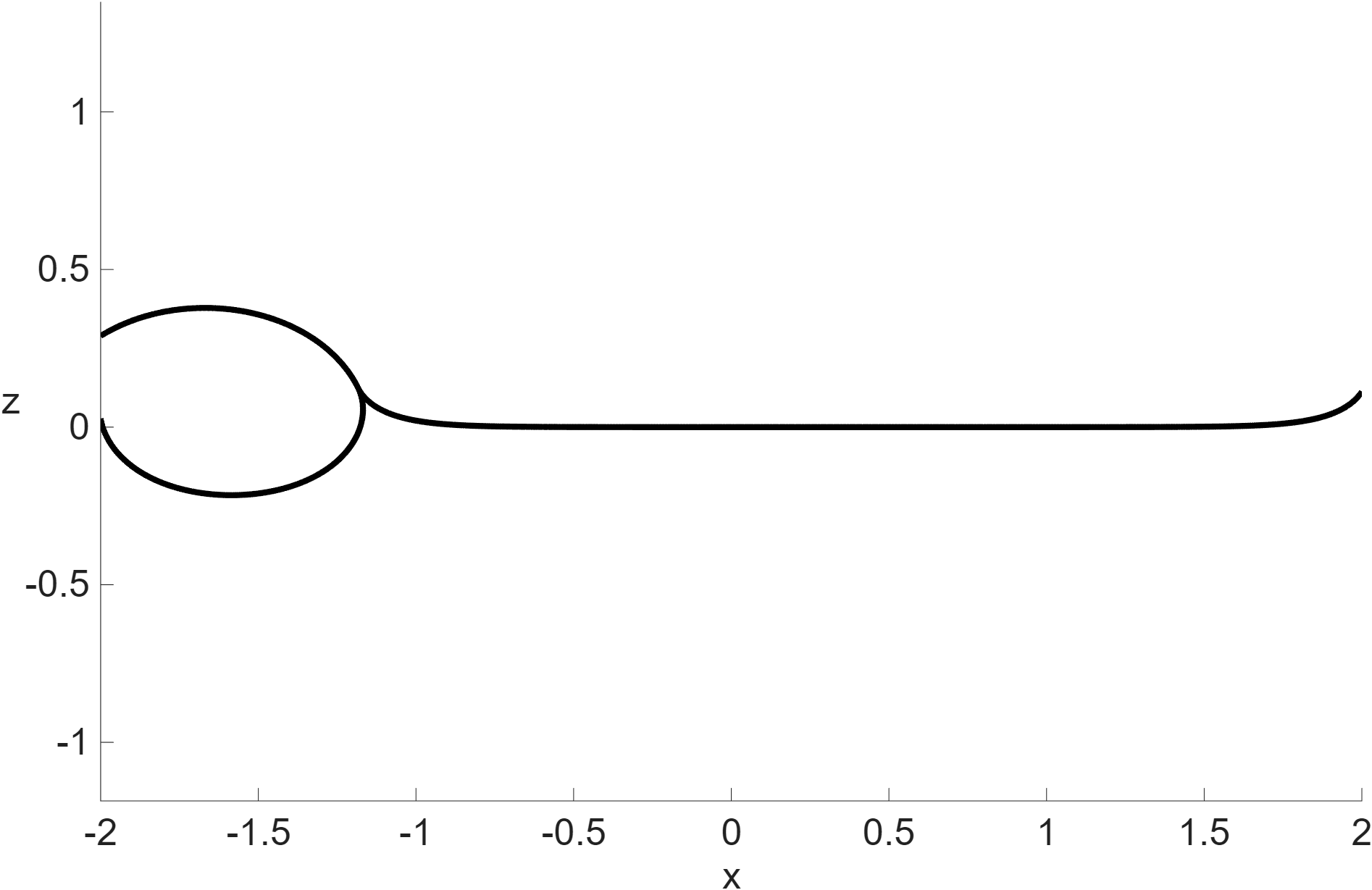}
	\qquad
	\includegraphics[scale=.2]{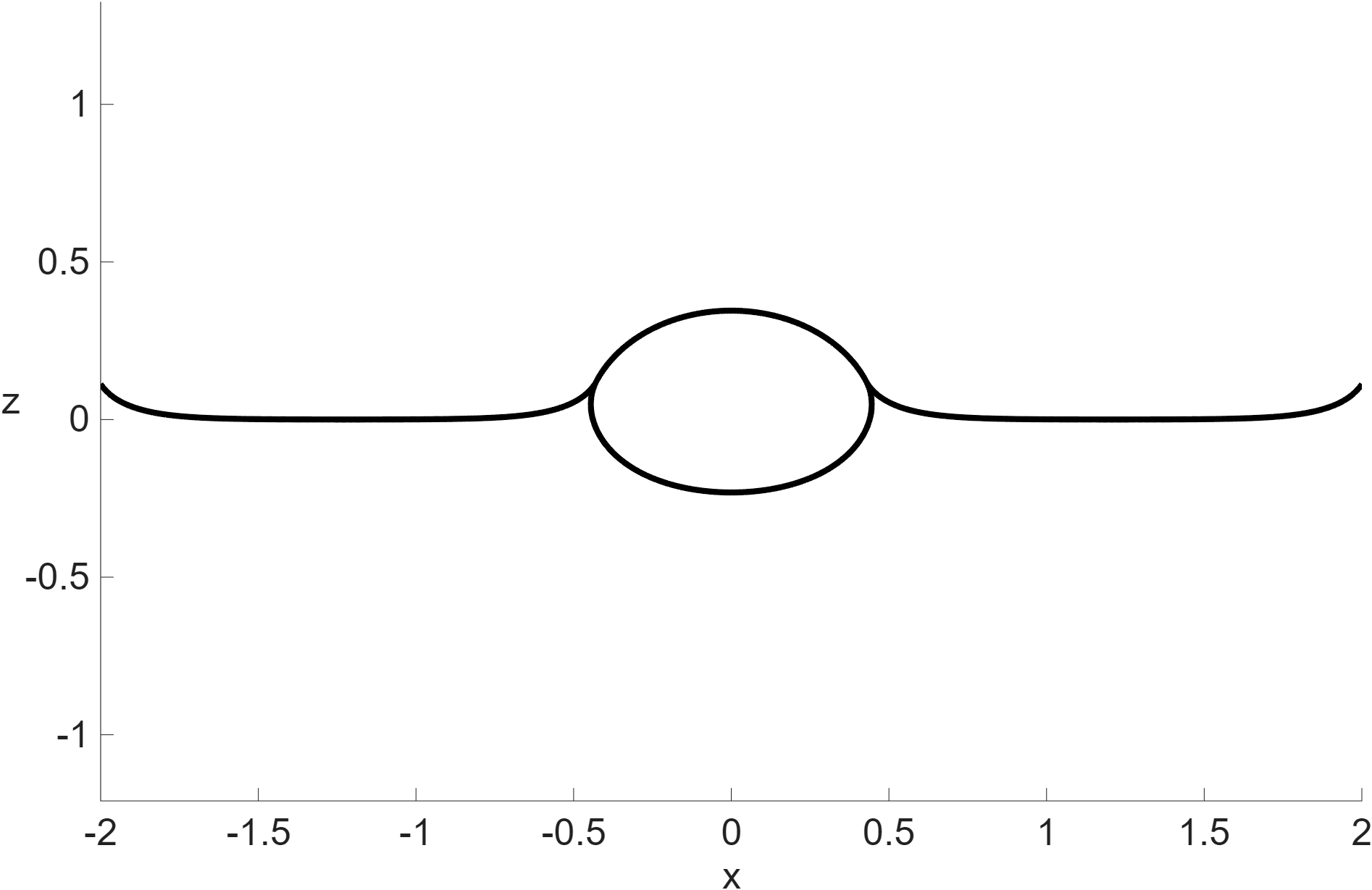}
	\caption{The wall-bound drop displayed here has lower energy than the displayed centrally located drop.  Here $\gamma^2_{0p} = 0.56$ is near the left edge of the experiment in Figure~\ref{fig:2dantiheuristic}.}
	\label{fig:2dantiheurisiticExamples}
\end{figure}

\begin{figure}[h]
	\centering
	\includegraphics[scale=.2]{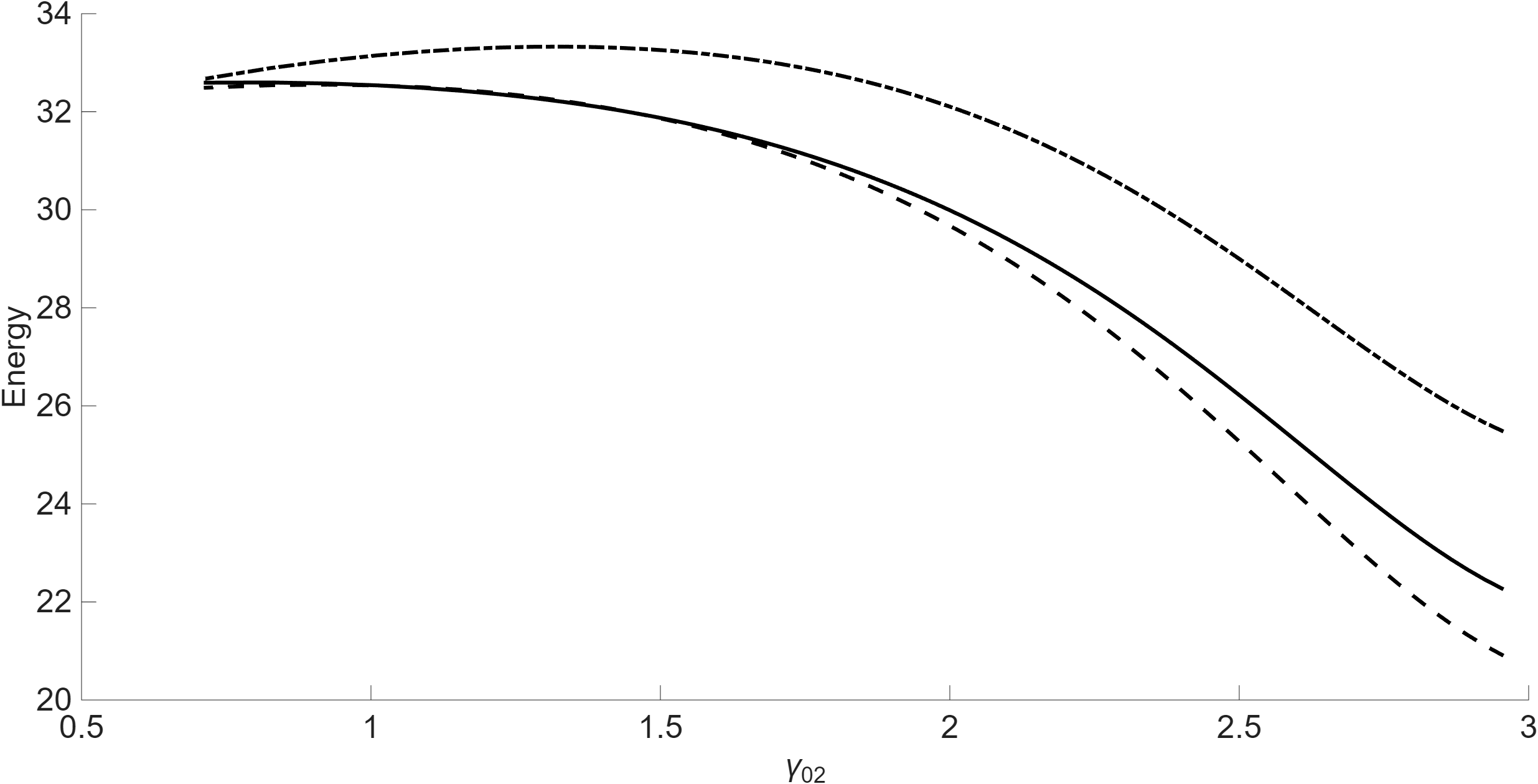}
	\caption{Here is an experiment where $\gamma_{02}$ is varied over a range of values.  The solid curve is the energy of the centrally located drop, the dashed curve is for the wall-bound drop, and the dot-dashed curve is for the drop split evenly on each wall.  The physical parameters that determine these are a drop volume of 0.4, a tube radius $X = 2$, densities $\rho_1 = 7.5$ and $\rho_2 = 15$, plate angles $\gamma^2_{0p} = \pi/2$ and $\gamma^1_{2p} = 1.252$, and the surface tensions range over the following values in this experiment as $\sigma_{01} \in[2.2,7.8], \sigma_{02}\in[2.2,7.8]$, and $\sigma_{12} = 6$. }
	\label{fig:2dgamma02}
\end{figure}

\begin{figure}[h!]
	\centering
	\includegraphics[scale=.2]{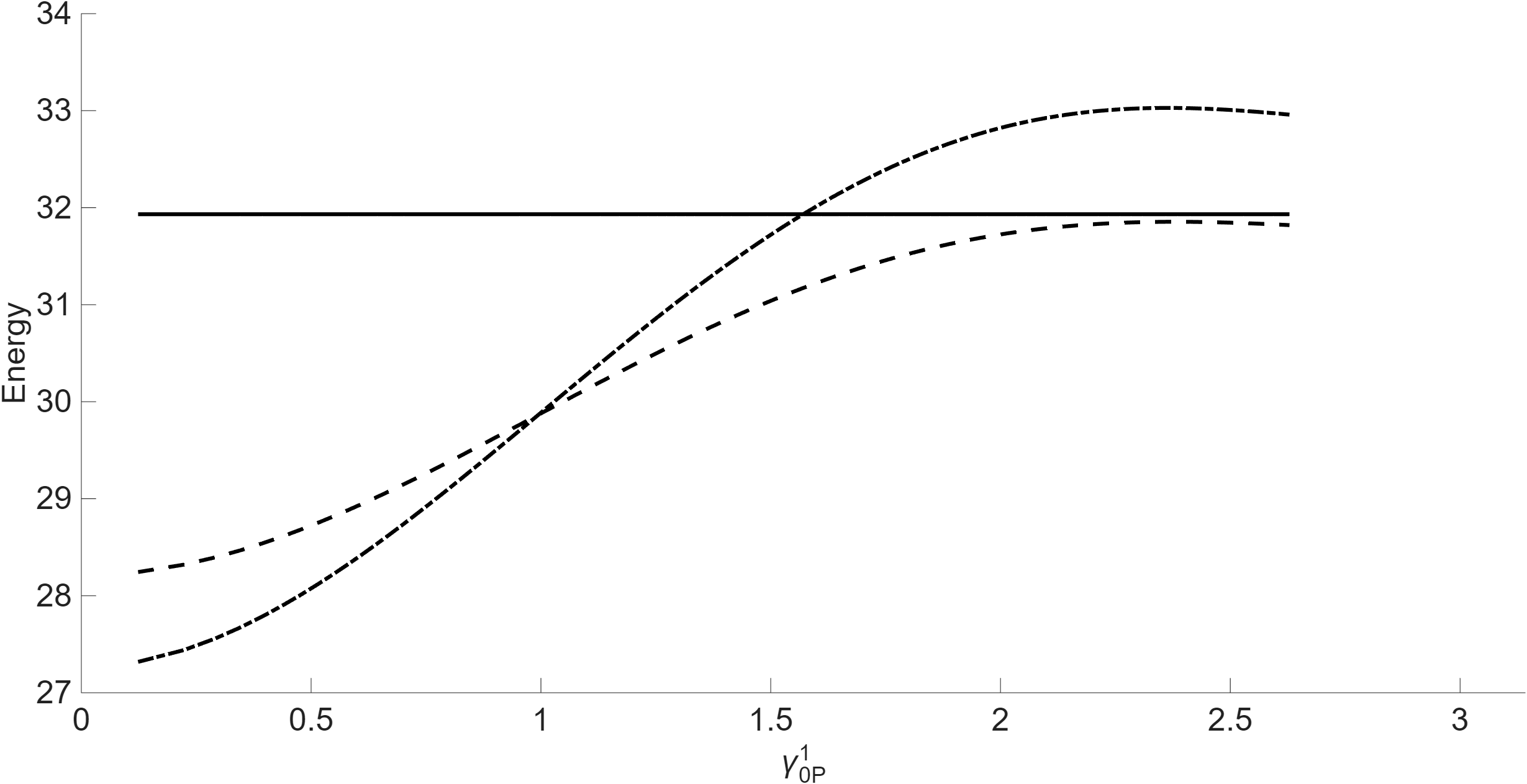}
	\caption{Here is an experiment where $\gamma^1_{0p}$ is varied over a range of values.  The solid curve is the energy of the centrally located drop, the dashed curve is for the wall-bound drop, and the dot-dashed curve is for the drop split evenly on each wall.  The physical parameters that determine these are a total volume of 0.4, a tube radius of $X = 2$, densities of $\rho_1 = 1$ and $\rho_2 = 15$, surface tensions $\sigma_{01} = 3, \sigma_{02} = 7$, and $\sigma_{12} = 6$, and the plate angle $\gamma^2_{0p}$ is fixed at $\pi/2$ while $\gamma^2_{1p}$ is in a range of $[1.12,2.09]$, generating the displayed third plate angle.}
	\label{fig:2dplate}
\end{figure}

\subsection{More parameter space explorations}
\label{sec:2Dmore}

First, we return to the heuristic that we explored in the last section.  There we showed the generally typical behavior that if there is no drop present in the capillary tube and the surface is concave up, then adding a floating drop results in a centrally located drop obtaining lower energy than the wall-bound drop.  In Figure~\ref{fig:2dantiheuristic} we show a parameter study where $\gamma^2_{0p}$ is given over a range of values.  The usual phenomena occurs where the energy profiles cross at a value of $\pi/2$, but on the left side of the figure the curves cross an additional time.  This gives a counter example to the heuristic, and Figure~\ref{fig:2dantiheurisiticExamples} shows drops from the unexpected region on the left.  It should be pointed out that values of $\gamma^2_{0p}$ less than those shown in Figure~\ref{fig:2dantiheuristic} are not possible, as the governing equation \eqref{eqn:plate} is not solvable there.  While our heuristic might be a generally useful guiding intuition, the interplay between the physical parameters is more subtle than this simple parameter test suggests.  We have not found a sharp criterion that indicates when such an unexpected phenomena might occur.

In our next example, we consider a variation in the surface tensions.  We exhibit this by taking a range of $\gamma_{02}$, which is the contact angle on the interior of the drop.  As that angle goes to zero, and generally without ever achieving 0, the fixed-volume floating drop problem will limit to a layer of the drop covering the entire tube diameter.  In Figure~\ref{fig:2dgamma02} we show energy profiles for centrally located drops, wall-bound drops, and evenly spit volumes of wall bound drops attached to each wall.

We end this subsection with an example where another plate angle is varied.  With $\gamma^2_{0p} = \pi/2$ fixed, $\gamma^1_{0p}$ is varied, and this is shown in  Figure~\ref{fig:2dplate}.   Here the drop split evenly against each wall can be seen to be the minimizer for a range of $\gamma^1_{0p}$ on the left.

\subsection{Symmetry and asymmetry}
\label{sec:asymmetry}

The most important question we address for floating drops in $\mathbb{R}^2$ is that of symmetry.  In this setting we have the tools at hand to compute centrally located drops, drops bound to a single wall, and drops split and adjacent to both walls.  The configurations that we do not discuss are off-center drops that do not touch a wall as well as split drops that do not touch a wall.  Reflection arguments make the off-center drops unlikely to be energy minimizing, and it is not clear that they lead to solvable Euler-Lagrange equations.  Split drops that do not meet the wall, and versions of those configurations with multiple drop components not meeting the wall also seem unlikely to be energy minimizing and also may not have solvable Euler-Lagrange equations.  So we consider four types of configurations, where our split drops could be evenly or unevenly divided.  

\begin{figure}[t]
	\centering
	\includegraphics[scale=.2]{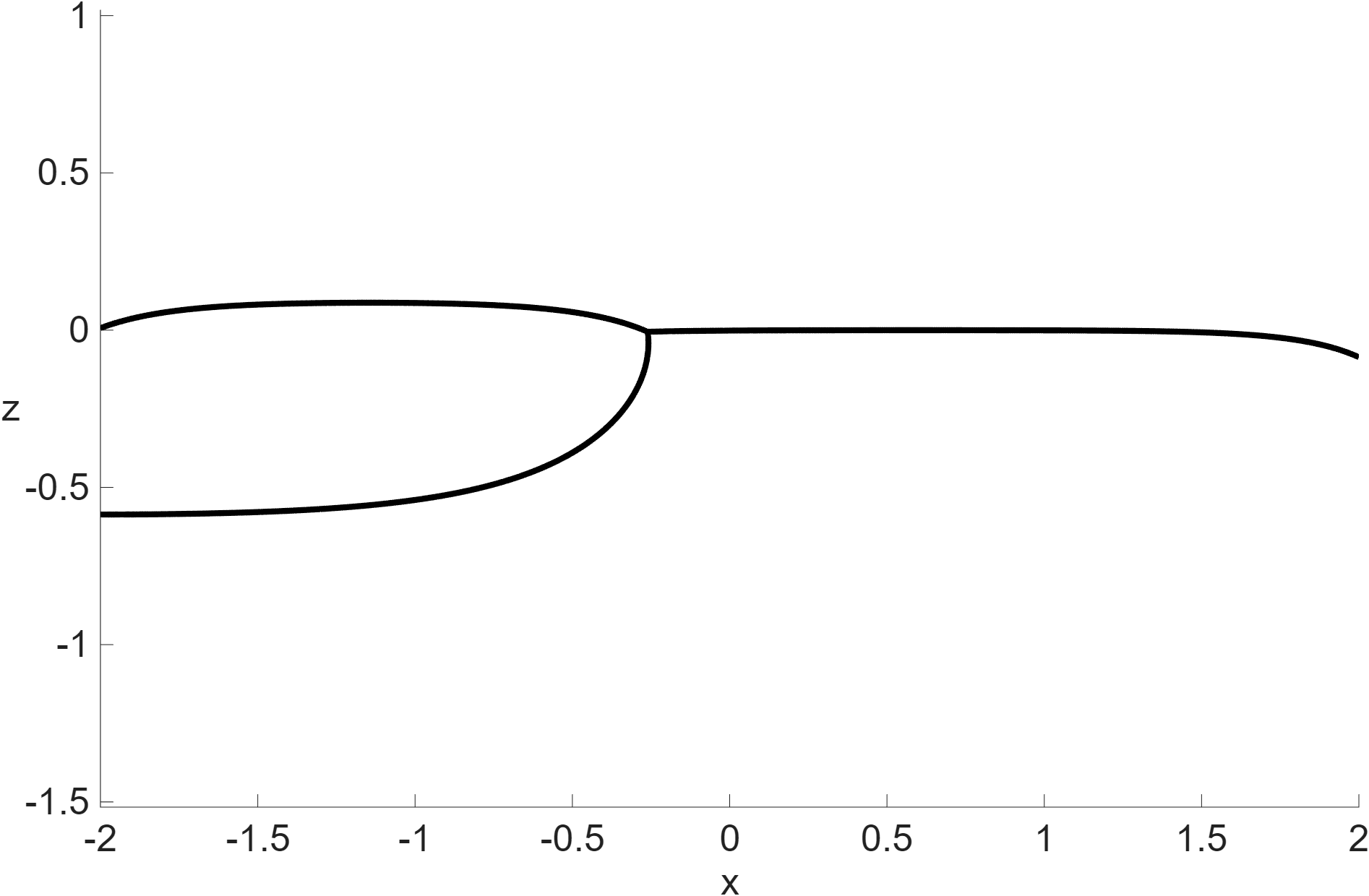}
	\qquad
	\includegraphics[scale=.2]{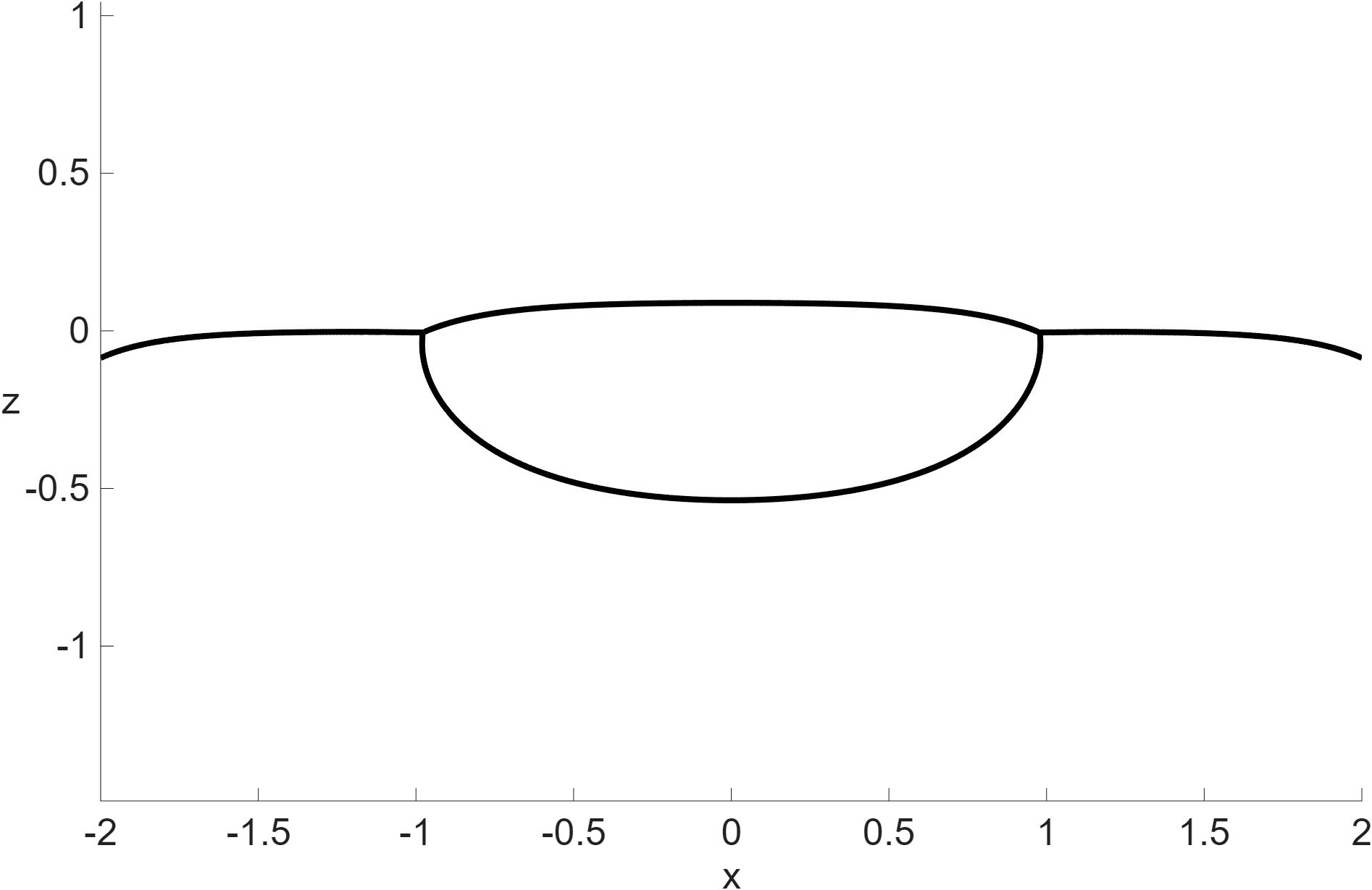}
	\caption{Floating drop configurations with  a volume of 1, tube radius of $X=2$, densities $\rho_1 = 13$ and $\rho_2 = 15$, surface tensions $\sigma_{01} = 7, \sigma_{02} = 6$, and $\sigma_{12} = 3$, and plate angles 
		$\gamma^2_{0p} = 2$, and $\gamma^2_{1p} = \pi/2$.  On the left is a drop adjacent to the left wall and this is the energy minimizer.  On the right is a centrally located drop. }
	\label{fig:2dWallAndCentralDrops}
\end{figure}

\begin{figure}[t]
	\centering
	\includegraphics[scale=.2]{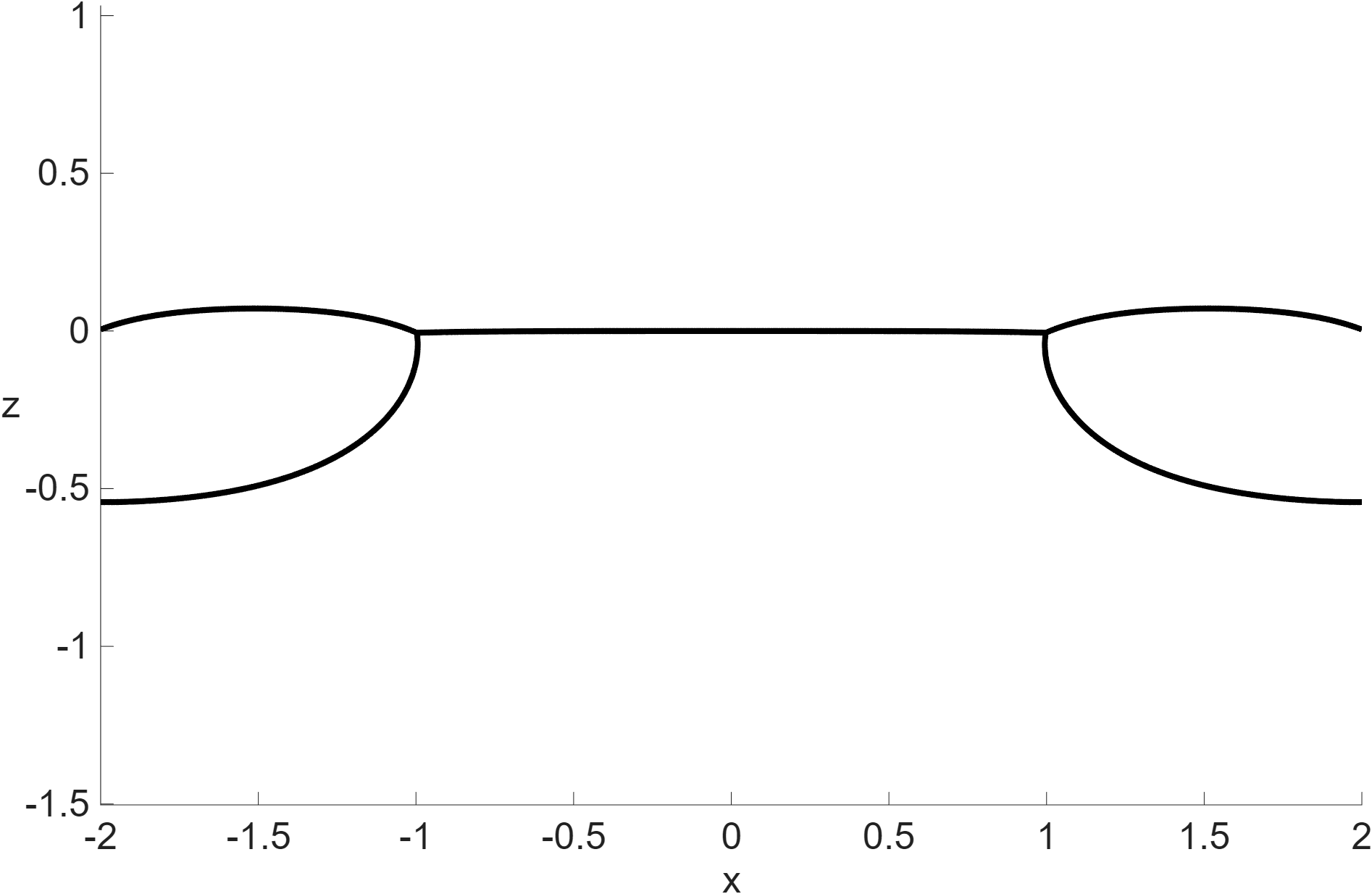}
	\qquad
	\includegraphics[scale=.2]{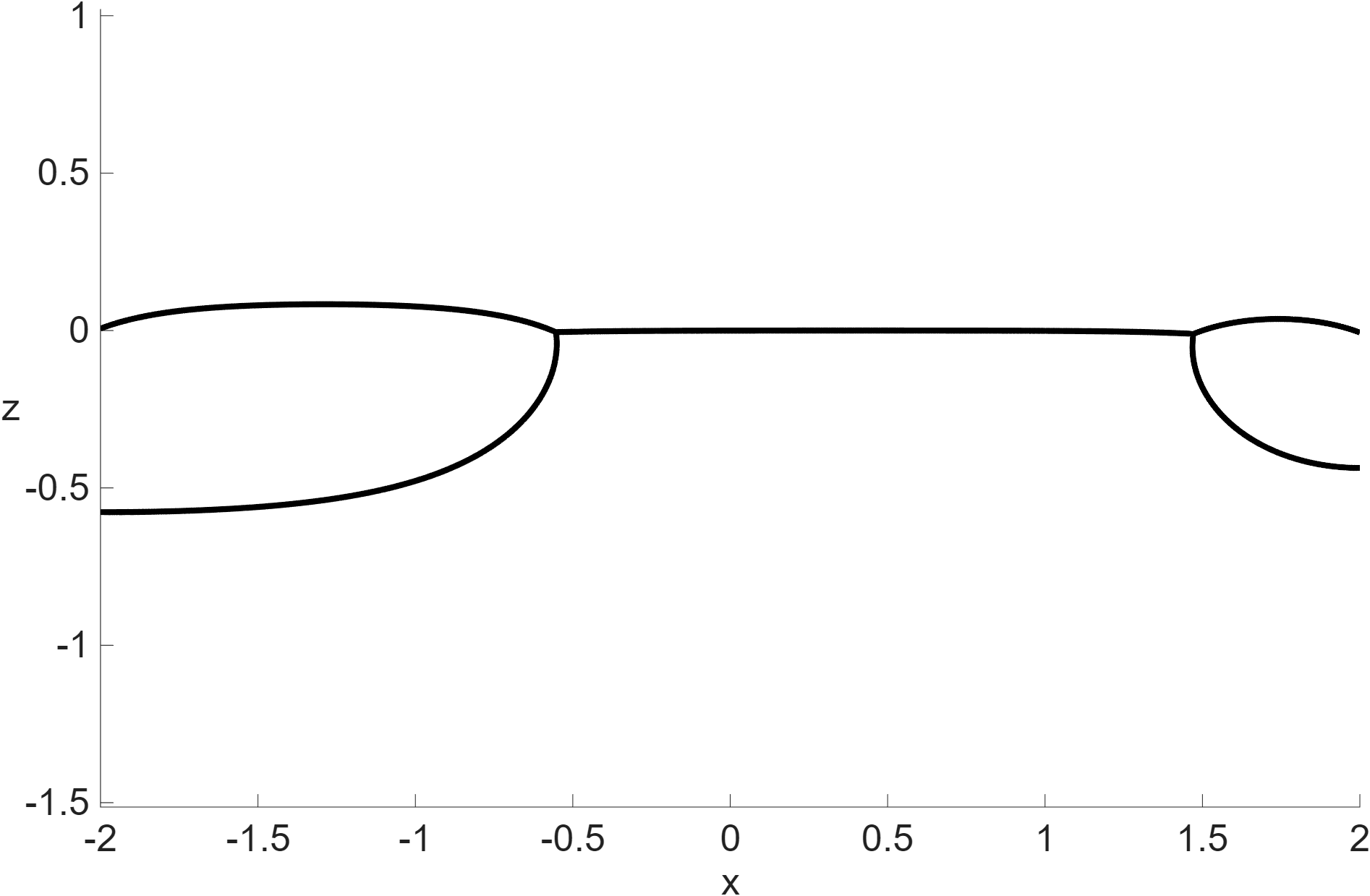}
	\caption{Floating drop configurations with the same physical parameters as Figure~\ref{fig:2dWallAndCentralDrops}.  On the left is a drop adjacent to the left wall, and this is the energy minimizer.  On the left is a drop evenly split between two areas attached to each wall, and on the right is an unevenly split drop with 1/5th the area adjacent to the right wall.}
	\label{fig:2dTwoSplitDrops}
\end{figure}

\begin{figure}[h]
	\centering
	\includegraphics[scale=.2]{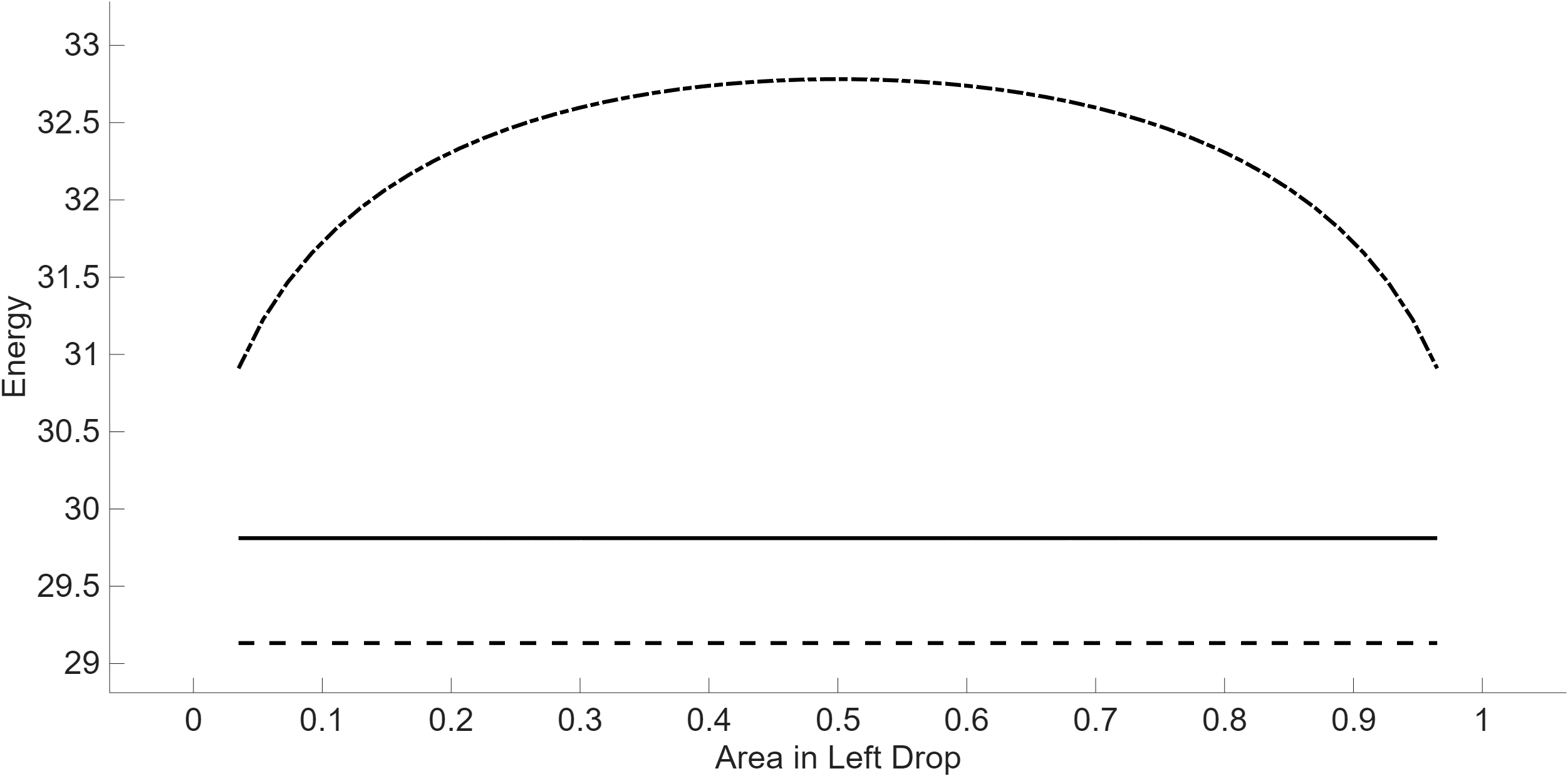}
	\caption{The energy of the split area drops with the portion of the area on the left side parameterizing the configurations.  The energy of the split area configurations are shown by a dot-dashed curve, and the solid (constant) curve is the energy of the centrally located drop.}
	\label{fig:2dSplitDropEnergyWallWins}
\end{figure}

In our first example, we consider a configuration where the single wall-bound drop is the energy minimizer.  This is shown in Figure~\ref{fig:2dWallAndCentralDrops}, where the centrally located drop is also displayed.  Then in Figure~\ref{fig:2dTwoSplitDrops} split drops are shown, with the evenly split drop appearing on the left, and an asymmetric split drop configuration on the right.  These asymmetrically split drops have never been observed to be energy minimizers, as one of the other three configurations shown has always produced a lower value of the potential energy.  Figure~\ref{fig:2dSplitDropEnergyWallWins} shows the energy profile as the volume is shifted from the right side of the tube to the left side of the tube.  It should be emphasized that this is an example of symmetry breaking in that the asymmetric floating drop configuration shown here is observed to be the energy minimizer.

\begin{figure}[t]
	\centering
	\includegraphics[scale=.2]{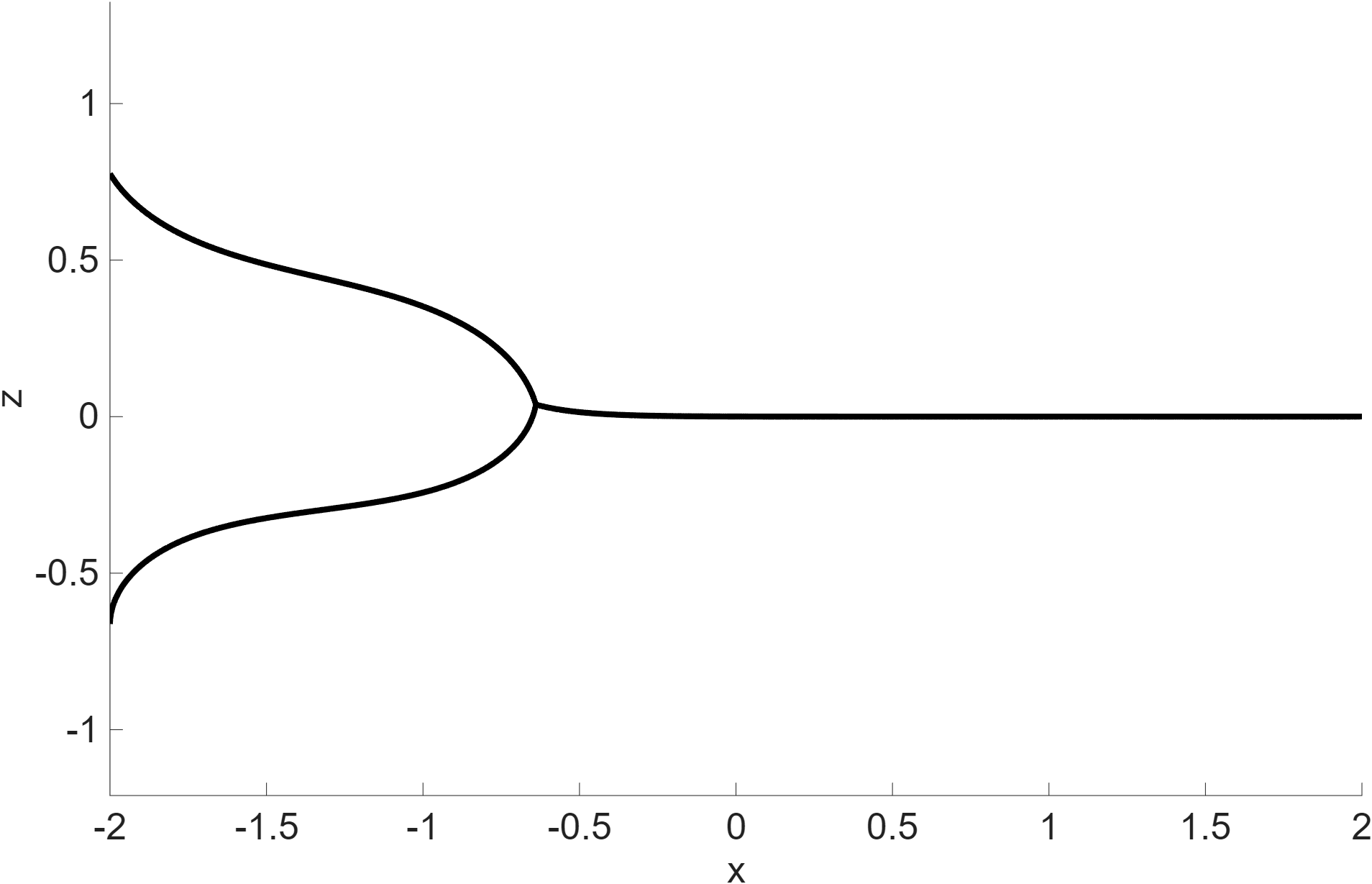}
	\qquad
	\includegraphics[scale=.2]{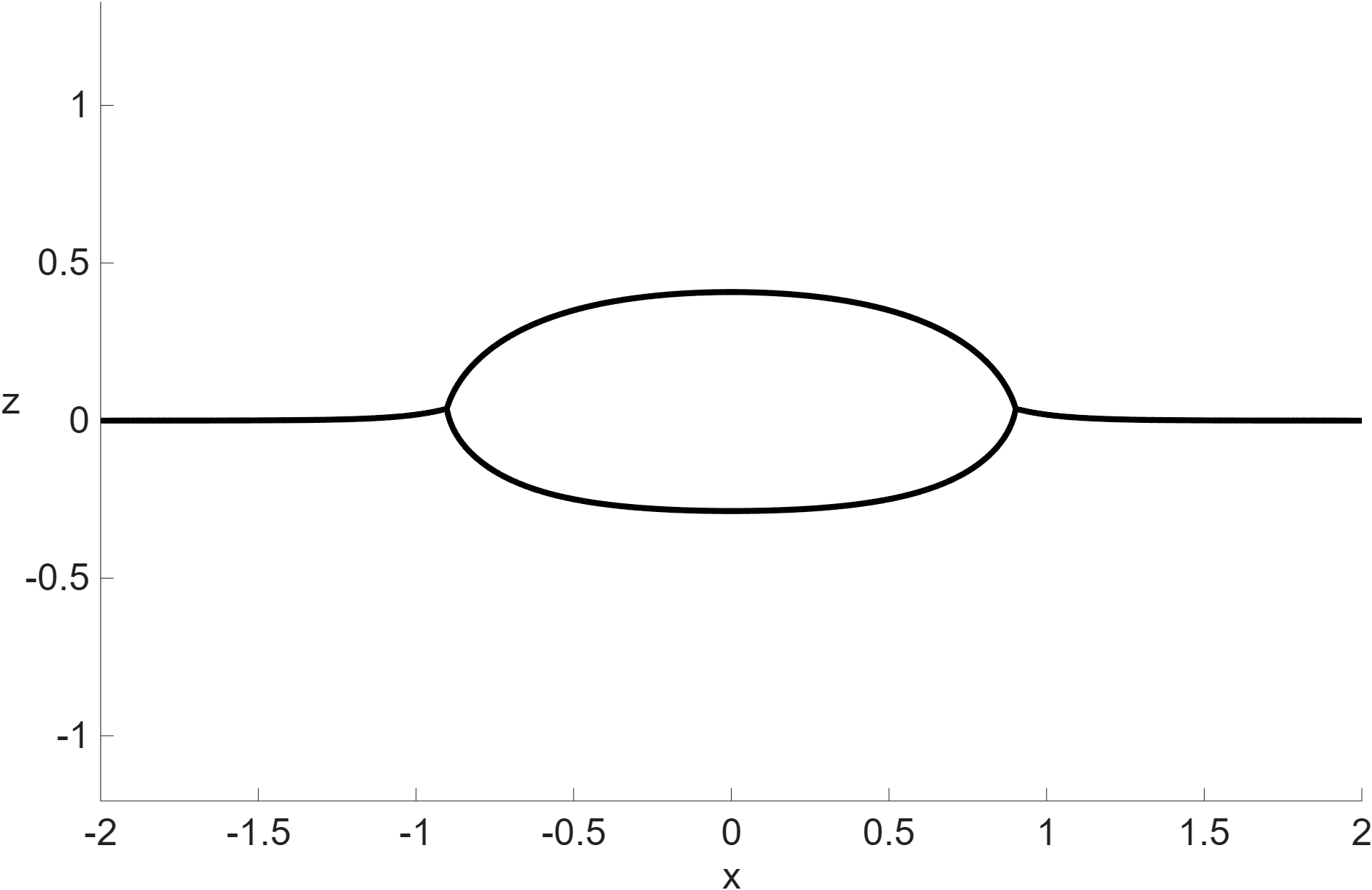}
	\caption{Floating drop configurations with an area of 1, tube radius of $X=2$, densities $\rho_1 = 6$ and  $\rho_2 = 15$, surface tensions $\sigma_{01} = 7, \sigma_{02} = 3$, and $\sigma_{12} = 6$, and plate angles 
		$\gamma^2_{0p} = \pi/2$, and $\gamma^2_{1p} = \pi - 0.01$.  On the left is a drop adjacent to the left wall.  On the right is a centrally located drop. }
	\label{fig:2dWallAndCentralDropsSplitWins}
\end{figure}

\begin{figure}[t]
	\centering
	\includegraphics[scale=.2]{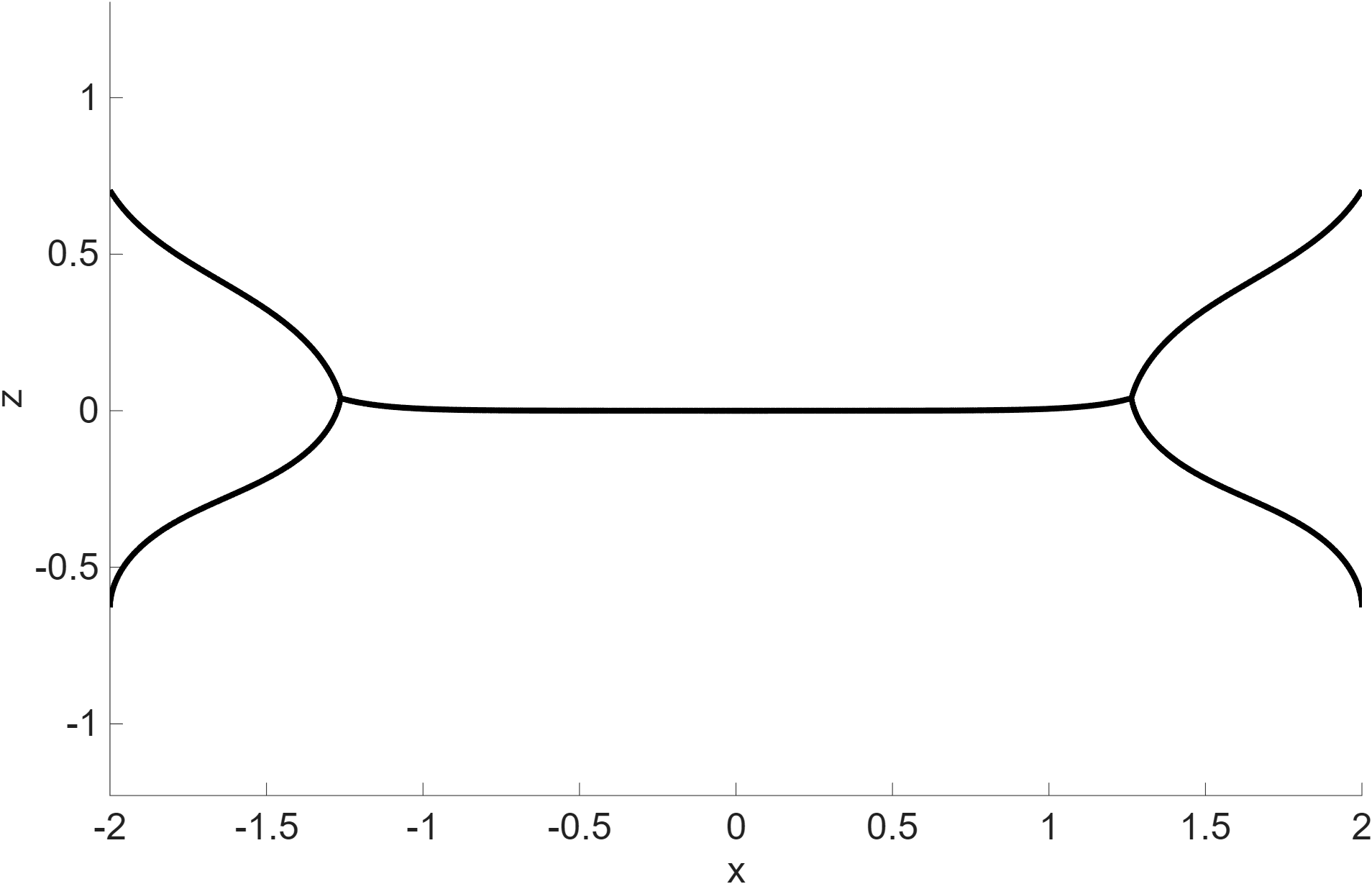}
	\qquad
	\includegraphics[scale=.2]{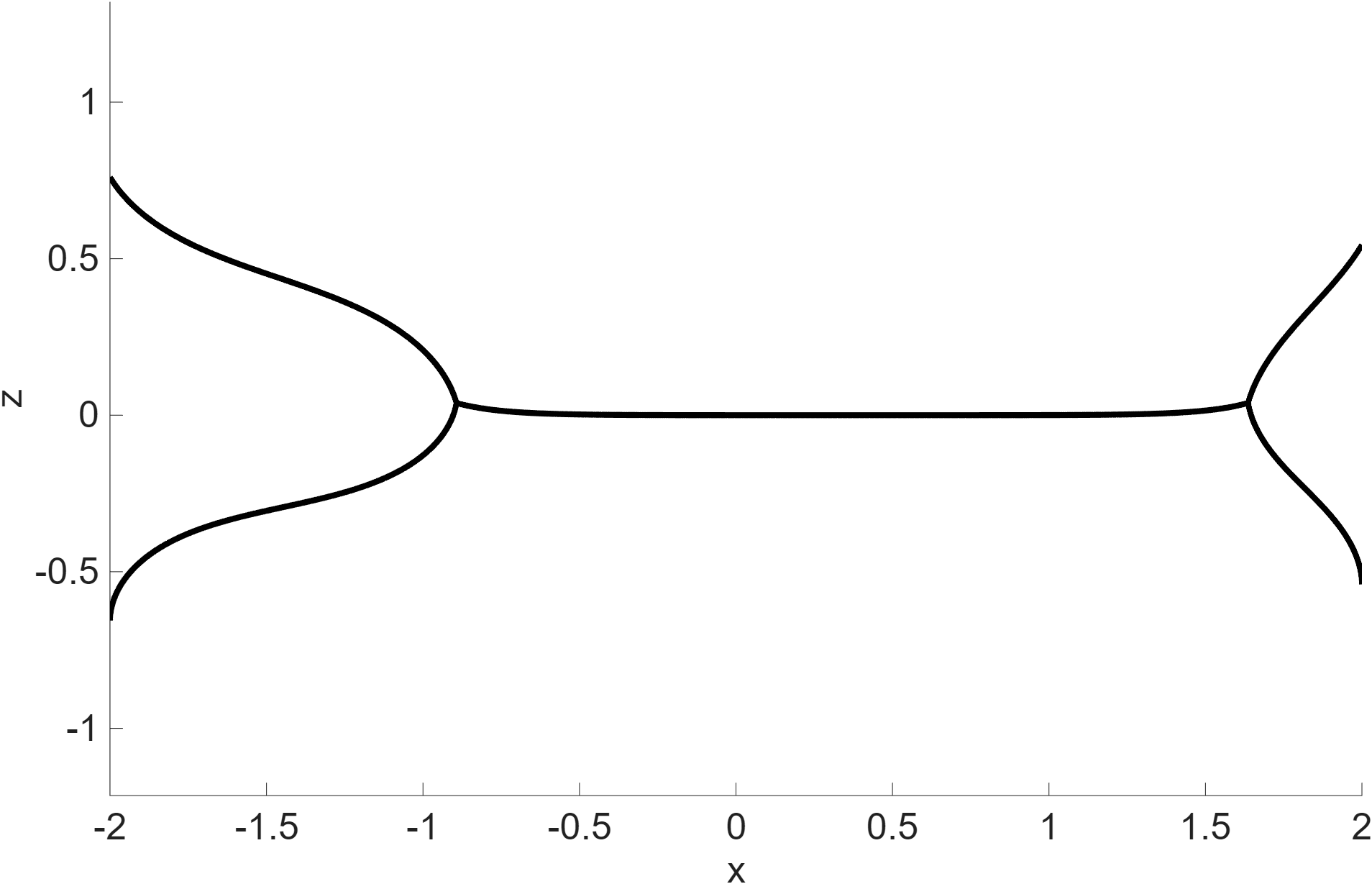}
	\caption{Floating drop configurations with the same physical parameters as Figure~\ref{fig:2dWallAndCentralDropsSplitWins}.  On the left is a drop evenly split between two areas attached to each wall, and this is the energy minimizer.  On the right is an unevenly split drop with 1/5th the area adjacent to the right wall. }
	\label{fig:2dTwoSplitDropsSplitWins}
\end{figure}

\begin{figure}[h]
	\centering
	\includegraphics[scale=.2]{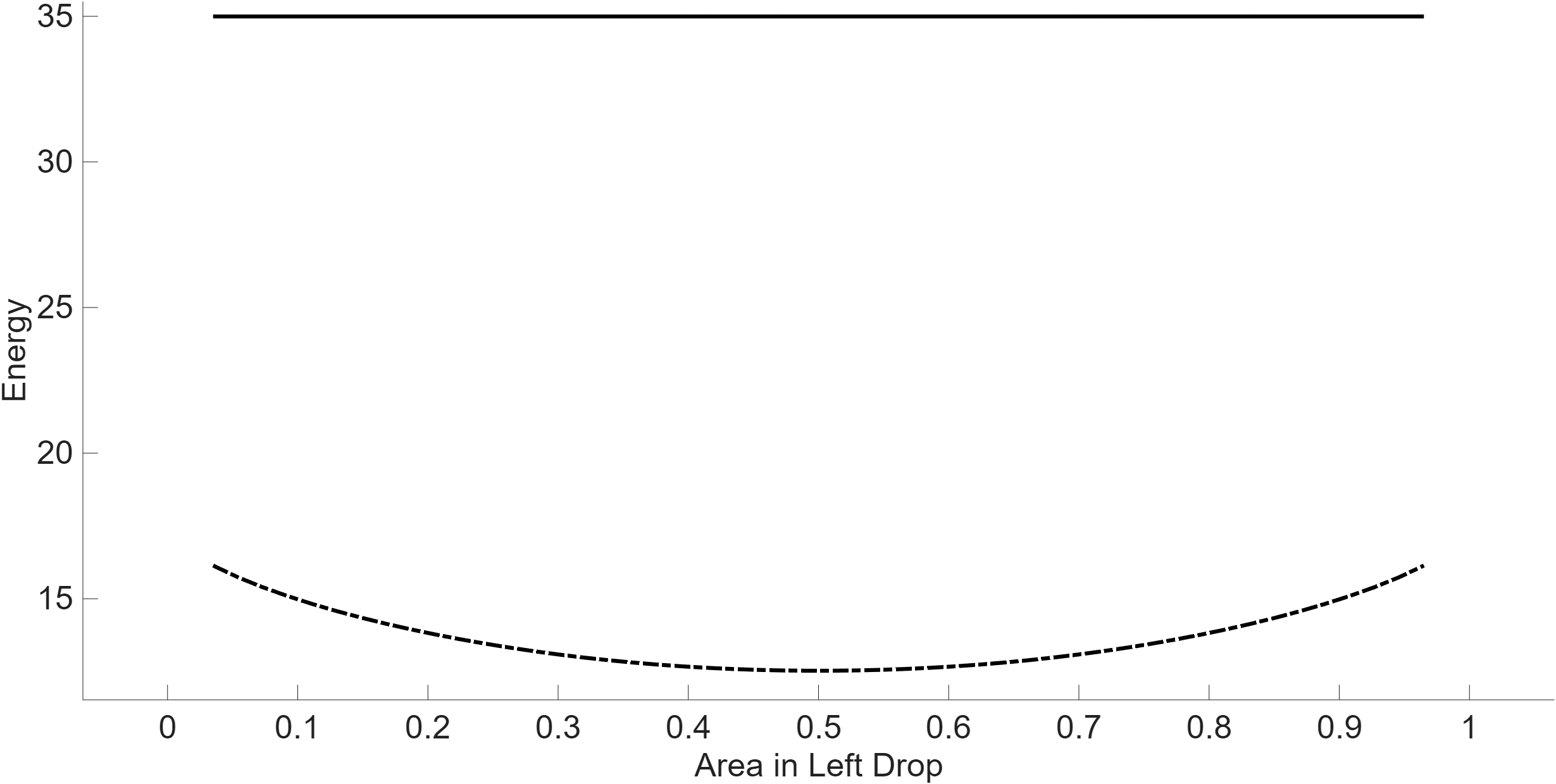}
	\caption{The energy of the split area drops with the portion of the area on the left side parameterizing the configurations.  The energy of the split area configurations are shown by a dot-dashed curve, and the solid (constant) curve is the energy of the centrally located drop.}
	\label{fig:2dSplitDropEnergySplitWins}
\end{figure}

In our second example, we consider a configuration where the evenly split wall-bound drop is the energy minimizer.   Figure~\ref{fig:2dWallAndCentralDropsSplitWins} shows a wall bound drop and a centrally located drop.  Then in Figure~\ref{fig:2dTwoSplitDropsSplitWins} split drops are shown, with the evenly split drop appearing on the left, and an asymmetric split drop configuration on the right.   Figure~\ref{fig:2dSplitDropEnergySplitWins} shows the energy profile as the area is shifted from the right side of the tube to the left side of the tube.

\subsection{Examples of multiple equal-energy configurations}

\begin{figure}[h]
	\centering
	\includegraphics[scale=.2]{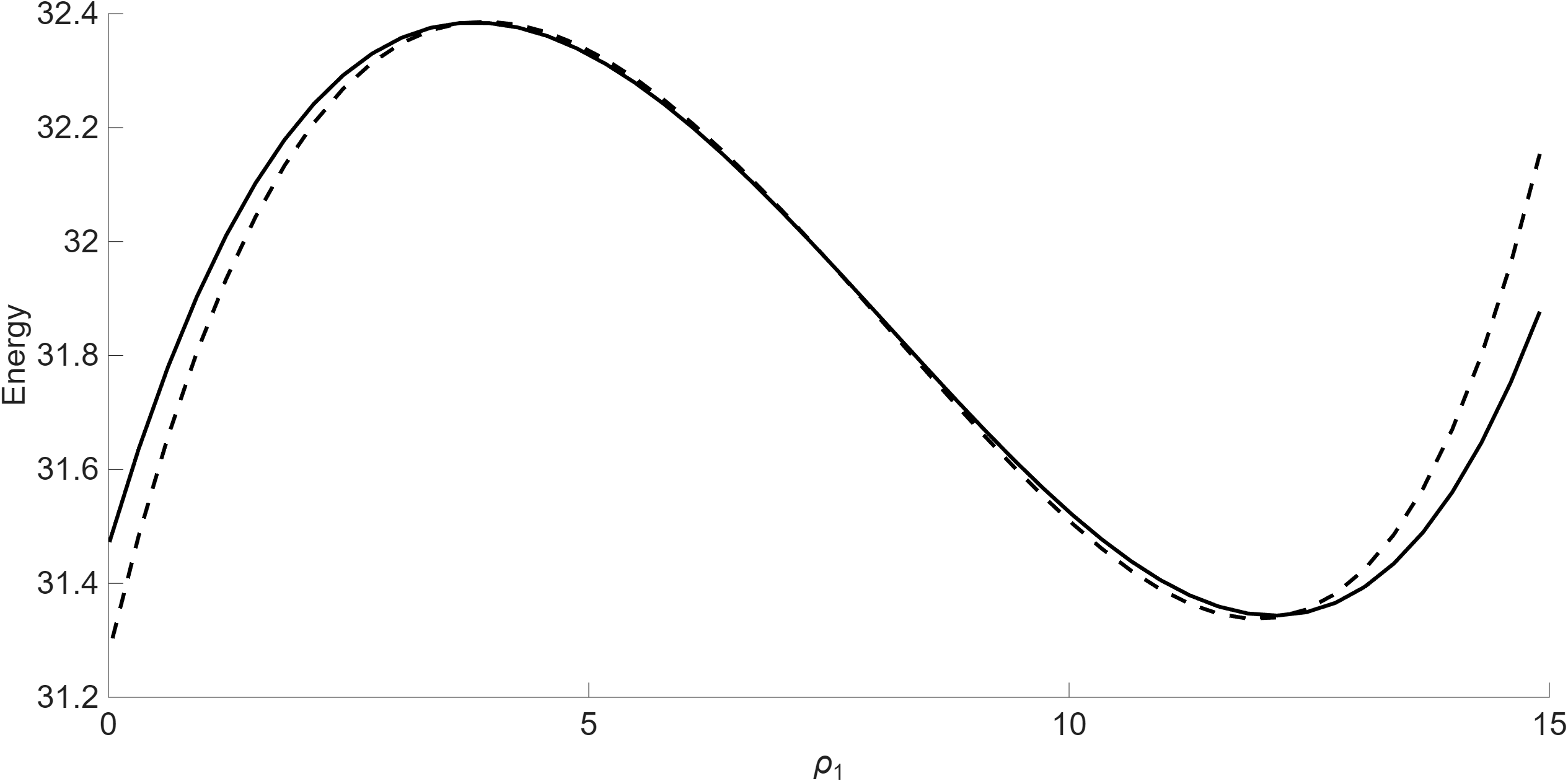}
	\caption{Comparing the energies of the centrally located drop (solid) and the wall-bound drop (dashed).  We are varying the density $\rho_1$.  There are three intersections of these curves in the displayed range.  The physical parameters are an area of 0.4 in a tube of radius $X = 2$, densities $\rho_1 = 7.5$ and $\rho_2 = 15$, surface tensions $\sigma_{01} = 3, \sigma_{02} = 7$, and $\sigma_{12} = 6$, and plate angles $\gamma^2_{0p} = \pi/2$ and $\gamma^2_{1p} = 1.252$}
	\label{fig:2dCaseCDensity}
\end{figure}

\begin{figure}[h]
	\centering
	\includegraphics[scale=.2]{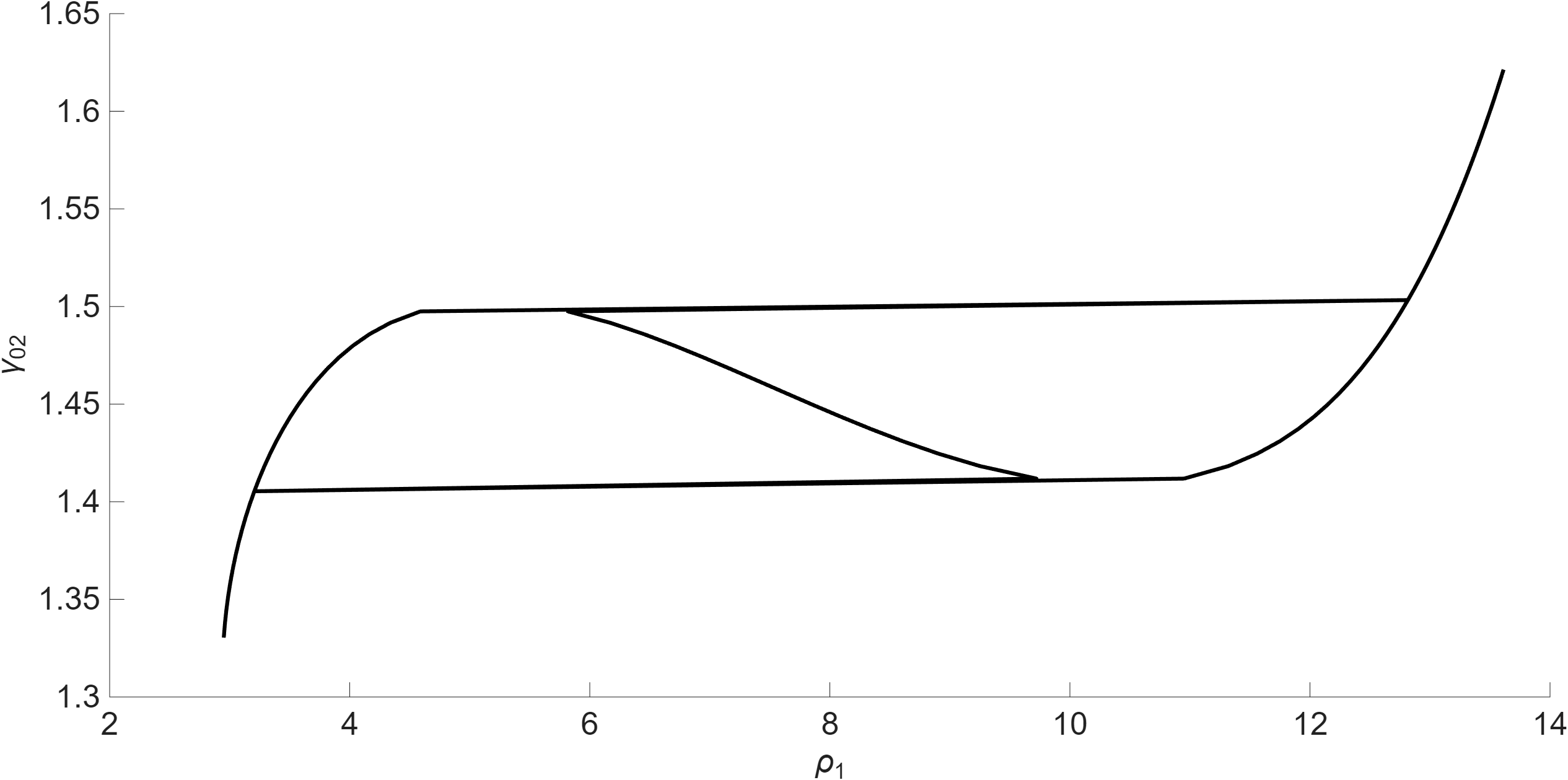}
	\caption{An overview of parameter studies where the number of configurations with equal energy is highlighted.  We consider a range of both $\rho_1$ and $\gamma_{02}$, and the curves represent values of $(\rho_1,\gamma_{02})$ where the energy of the wall-bound drop is equal to the energy of the centrally located drop.   The physical parameters here are an area of 0.4, a tube radius of $X = 2$, density $\rho_2 = 15$,  plate angles $\gamma^2_{0p} = \pi/2$ and $\gamma^1_{2p} = 1.252$,  and the surface tension $\sigma_{12} = 6$, while $\sigma_{01}$ ranges over $[2.8, 3.2]$ to produce the range of $\gamma_{02}$ displayed.}
	\label{fig:2ddesityFlow}
\end{figure}

Finally, we show in Figure~\ref{fig:2dCaseCDensity} a density experiment where there are three values of $\rho_1$ where the wall-bound drop and the centrally located drop have the same energy value.  We are able to find many cases with this number of equal-energy examples, and in Figure~\ref{fig:2ddesityFlow} we show a range of two parameters where the curves trace out trajectories where the energies are equal.  This shows that for some values of $\rho_1$ there are one, two, or three values of $\gamma_{02}$ that have equal energy values for the centrally located drops and the wall bound drops.  It would be possible to run these experiments for different choices of pairs of parameters.

\section{Validation of the numerical method}
\label{Zero}

Finally, we build a closed-form solution to the floating drop problem in the absence of gravity and then we show that our numerical method can approximate that solution quite accurately, even though our solver will not directly treat the zero gravity case.  This was requested by one of the anonymous referees to validate our numerical method, and the example we give had been suggested to us by John McCuan in the Fall of 2025. 

We work in the lower dimensional model for ease of constructing the following explicit solution.  We consider the case where all of the surface tensions are equal: $\sigma_{01} = \sigma_{02} = \sigma_{12} = \sigma$ for some constant $\sigma > 0$ and where the wetting energies are homogeneous, so that the interfaces are orthogonal to any walls they may meet.  We will consider the tube width to be the interval $[0,X]$, for some fixed $X>0$.

\subsection{A drop floating without gravity}

While the techniques used are elementary, these results mirror the features seen previously, and give the first rigorous proof of the non-uniqueness of solutions to the free boundary problems under discussion.

If we set $g=0$, then the formulation we have presented in this paper can be applied with some small exceptions.  The underlying differential equation
\begin{equation}
	H = -\lambda
\end{equation}
may have solutions that satisfy the boundary conditions 
\begin{eqnarray}
	\psi = \sin\psi_a, \quad r = a, \\
	\psi = \sin\psi_b, \quad r = b. 
\end{eqnarray}
However, these solutions are only unique up to vertical translations.  Due to this, we proceed to construct a floating drop directly.  

We first observe that the Lagrange multiplier is zero for the interface $S_{02}$, so it has a constant height and we may assume that $w \equiv 0$.  This anchors the problem with respect to the arbitrary vertical translations.  This also implies that the inclination angle of $S_{02}$ is zero at the free boundary.  Next, we observe that for $S_{01}$ and $S_{12}$, the Lagrange multipliers are opposite of each other.  Since these curvatures are constant, $S_{01}$ and $S_{12}$ are circular arcs with the same radius of curvature.   

The free boundary conditions \eqref{neumann_triangle} on the interior contact angles imply that 
\begin{equation}
	\gamma_{01} = \gamma_{02} = \gamma_{12} = \frac{2\pi}{3}.
\end{equation}

We will reduce to four main cases of problems: trivial drops, the wall bound drop, a family of split area drops where the drop is adjacent to each wall, and loosely speaking, the centrally located drop.  Since $w\equiv 0$, this last case need not be restricted to a drop being in the center of the tube, but describes the one-parameter family of solutions where the drop slides left or right without touching either wall, all of which have the same energy.  For simplicity we will refer to this last case as the centrally located drop.

\begin{figure}[h!]
	\centering
	\includegraphics[scale=.3]{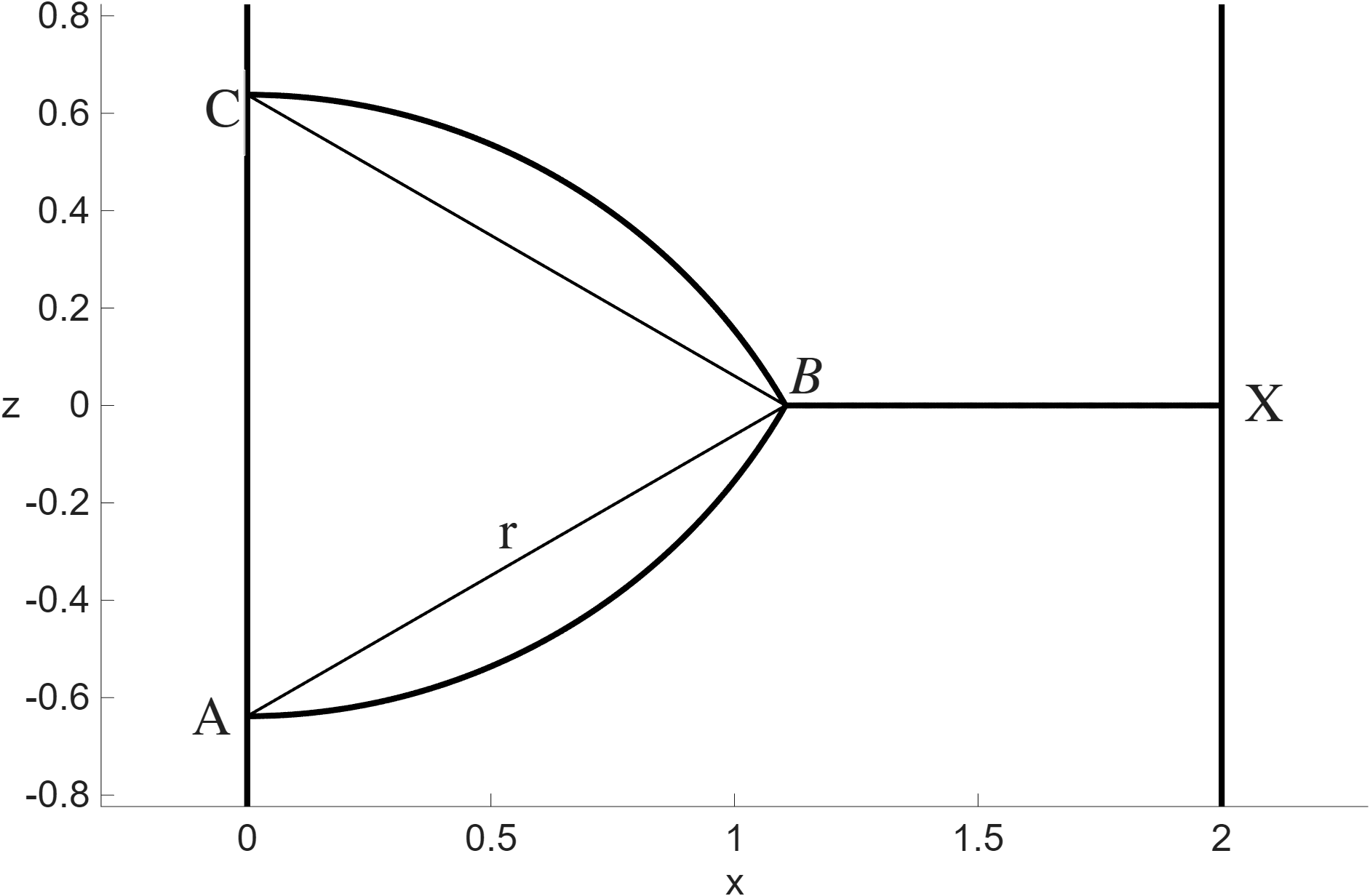}
	\caption{A schematic for the zero gravity floating drop.}
	\label{fig:lowGrav2}
\end{figure}

Furthermore, as the interfaces have inclination angle zero at the left boundary wall, the centers of the circular arcs describing $S_{01}$ and $S_{12}$ are both on that vertical line $x = 0$.  See Figure~\ref{fig:lowGrav2}.  To describe the geometry, we define the intersection of $S_{12}$ with $x = 0$ to be the point $A$, the triple junction point to be $B $, and the intersection of $S_{01}$ with $x = 0$ to be the point $C$.  Then if $E$ is the center of the circular arc describing $S_{01}$, $E = (0,-h)$ for some quantity $h$.  As usual, we denote the origin by $O$.  Then, as the radius $EB$ is orthogonal to the circular arc $S_{01}$, and the angle $\gamma_{12} = 2\pi/3$, the angle $\angle OBE = \pi/6$, and this with the fact that the radius of curvature $r = |EC|$ implies that $E=A$.  Then either by symmetry, or by repeating the argument, we also find that the center of the circular arc describing $S_{12}$ is  $C$.  It immediately follows that $\bigtriangleup ABC$ is an equilateral triangle with side length $r$ and interior angles $\pi/3$.  Then the enclosed area of the drop is given by adding the areas of the two circular sectors and subtracting the doubly covered triangle area.  That is,
\begin{equation}
	\label{eqn:radii}
	A = \left( \frac{\pi}{3} - \frac{\sqrt{3}}{4}\right) r^2.
\end{equation}
It will be helpful to define the constant
\begin{equation}
	C := \frac{\pi}{3} - \frac{\sqrt{3}}{4} > 0.
\end{equation}
Of course, \eqref{eqn:radii} could easily be solved for $r = r(A)$ and we could deduce the energy quantities directly from the area $A$, but there is a cleaner approach.  We will note that the free boundary is given by 
\begin{equation}
	\bar x = \frac{\sqrt{3}r}{2}.
\end{equation}

Thus, for a single wall bound drop, the energy is 
\begin{eqnarray}
	\mathcal{E} &=& \sigma\left( \frac{2\pi}{3}r + X - \bar x \right) \nonumber\\
	&=& \sigma\left( 2Cr + X \right).
\end{eqnarray}

Then for the centrally located drop, we can construct the configuration by taking the center of the drop to be on the vertical axis, and build the right half of the configuration using the above arguments for a prescribed area $\frac{1}{2}A$ and a half-sized container $[0,\frac{1}{2}X]$, then reflect this configuration symmetrically about the vertical axis to obtain the full drop area and container size.  The resulting radius of curvature is given by 
\begin{equation}
	\frac{1}{2}A = C r^2_c
\end{equation}
and the corresponding energy is
\begin{equation}
	\mathcal{E} = \sigma\left( 4Cr_c + X  \right).
\end{equation}

The next part is to develop the split area drops.  We denote the radius of curvature for the left area $A_1$ by $r_1$ and for the right area $A_2$ we have $r_2$.  Then we can repeat the above construction on the left and the right without difficulty and the resulting free boundary points are denoted by $\bar x_1$ and $\bar x_2$.  Given that the total area is prescribed to be $A$, we have
\begin{equation}
	A = C(r_1^2 + r_2^2), 
\end{equation}
and we can obtain a relation between $r_1$ and $r_2 = r_2(r_1)$ by observing that
\begin{equation}
	0 = \frac{dA}{dr_1} = C\left(2r_1 + 2r_2\frac{dr_2}{dr_1} \right), 
\end{equation}
and it follows that 
\begin{equation}
	\frac{dr_2}{dr_1} = \frac{-r_1}{r_2}.
\end{equation}

Then we find the energy of the split drop to be
\begin{equation}
	\mathcal{E}(r_1) = \sigma \left(2C(r_1 + r_2) + X \right).
\end{equation}
and we take the derivative with respect to $r_1$ to obtain
\begin{equation}
	\frac{d\mathcal{E}(r_1)}{dr_1} = 2\sigma C\left(1 - \frac{r_1}{r_2}\right)
\end{equation}
and at the critical point we find that $r_1 = r_2$.  Finally, if $r_1 = 0$, the entire area is on the right, and as $C > 0$, the energy is increasing there.  As there is only one critical point, it must be a local energy maximum and it follows that the wall bound drop has lower energy than any of the split area drops.

If we are to compare the energies of the wall bound drop with the centrally located drop, then by setting their respective areas equal, we find
\begin{equation}
	r = \sqrt{2} r_c,
\end{equation}
and applying that to the energy formulations above, since 
\begin{equation}
	r_c > \frac{\sqrt{2}}{2}r_c,
\end{equation}
the single wall bound drop has lower energy than the centrally located drop.  

In the next two configurations we eliminate any other possibilities.  First we will show that any two drops floating in a central manner without attaching to the wall have higher energy than a single centrally floating drop, and this will eliminate any configurations from consideration where there are multiple disconnected drops not attached to a wall, and finally we will show a hybrid configuration with some area of fluid attached to a wall and the remainder of the fluid floating unattached to a wall.

For the first of these configurations, we denote the area on the left by $A_3$ and the area on the right by $A_4$, similarly to what we did before.   Then the respective radii of curvature are defined by
\begin{eqnarray}
	A_3 &=& 2Cr_3^2, \\
	A_4 &=& 2Cr_4^2, 
\end{eqnarray}
and if the total area $A = A_3 + A_4$ is prescribed we can find $r_4 = r_4(r_3)$, and derive the differential equality as before to find
\begin{equation}
	\frac{dr_4}{dr_3} = -\frac{r_3}{r_4}
\end{equation}
again in this case.  Then the energy of this configuration is given by
\begin{equation}
	\mathcal{E}(r_3) = \sigma \left( 4C(r_3 + r_4) + X \right),
\end{equation}
and as before, taking the derivative leads to $r_3 = r_4$ at the local maximum, and a lower energy state is given by a single floating drop not attached to a wall, as opposed to two or more disconnected unattached drops.

Our final split area configuration to consider is the hybrid case with a portion of the area attached to a wall and the remainder in a single drop unattached to a wall.  Here we arbitrarily place the drop adjacent to the left wall, denote $A_5$ to be that wall bound drop on the left, and the remainder of the floating fluid is denoted by $A_6$.  Then the corresponding radii of curvatures are given by the relations
\begin{eqnarray}
	A_5 &=& Cr_5^2, \\
	A_6 &=& 2Cr_6^2, 
\end{eqnarray}
and if the total area $A = A_5 + A_6$ is prescribed we can find $r_6 = r_6(r_5)$, and derive the differential equality as before to find
\begin{equation}
	\frac{dr_6}{dr_5} = -\frac{r_5}{2r_6}.
\end{equation}
Then the energy is
\begin{equation}
	\mathcal{E}(r_5) = \sigma \left( 2C(r_5 + 2r_6) + X \right),
\end{equation}
and as before, taking the derivative leads to $r_5 = r_6$ at the local maximum, and a lower energy state is given by a single floating drop  attached to a wall, as opposed to the hybrid configuration under  consideration.

The last two configurations to consider are trivial.  We choose an area small enough that the drop does not span the entire tube width, and this area restriction is 
\begin{equation}
	A < \frac{4}{3}CX^2
\end{equation}
for the configuration with a single area attached to a wall.  Other formulas are easy to compute for the various configurations above, and we simply make sure that our drop area is smaller than any of these quantities.

Finally, we consider the case where the drop is not attached to the interface $S_{02}$.  This case is only possible in the zero gravity case, but we consider it for completeness.  The radius of curvature of a free-floating drop is given by 
\begin{equation}
	A = \pi r_0^2,
\end{equation}
and the corresponding energy is
\begin{equation}
	\mathcal{E} = \sigma \left( 2\sqrt{\pi A} + X \right).
\end{equation}
Since $C < \pi$, this configuration has more energy than the single area attached to one wall and intersecting $S_{02}$.  The radius of curvature of a free-floating drop away from $S_{02}$, but attached to a wall is given by 
\begin{equation}
	A = \frac{1}{2}\pi \tilde r_0^2,
\end{equation}
and the corresponding energy is
\begin{equation}
	\mathcal{E} = \sigma \left( \sqrt{2\pi A} + X \right).
\end{equation}
Since $C < \pi/2$, thus this last configuration clearly has more energy than the single area attached to one wall  and intersecting $S_{02}$.

Thus we have proven the following 
\begin{theorem}
	The zero gravity floating drop problem in 2D with equal surface tensions displays multiple examples of non-uniqueness of solutions to the Euler-Lagrange equations, with one family of drops floating unattached to a wall giving a one-parameter family of solutions with equal energy.  The global energy minimizing configuration is given by a single area $A$ of fluid attached to one wall with circular arcs defining the upper and lower drop boundaries, with the radius of curvature for the drop interfaces given by \eqref{eqn:radii}, while the surface between the two unbounded areas of fluids is given by a flat interface, which can be taken to have height identically zero.
\end{theorem}

If one considers the above derivation in $\mathbb{R}^3$, we immediately obtain
\begin{corollary}
	The zero gravity floating drop problem with equal surface tensions in $\mathbb{R}^3$ displays multiple examples of non-uniqueness of solutions to the Euler-Lagrange equations, with one family of drops floating unattached to a wall giving a one-parameter family of solutions with equal energy.  In these examples a single volume $V$ of fluid unattached to any wall with spherical caps defining the upper and lower drop boundaries, with the radius of curvature for the drop interfaces dictated by the enclosed volume by the relation
	\begin{equation}
		V = \frac{5\pi}{4}r^3,
	\end{equation}
	while the surface between the two unbounded volumes of fluids is given by a flat interface, which can be taken to be identically zero.  These configurations where the drop forms a lens made up of the two spherical caps and this lens is free to move about the flat exterior interface so long as there is no contact with the tube boundary at radius $R$.  The constant energy for these configurations is given by
	\begin{equation}
		\mathcal{E} = \sigma\left( \frac{5\pi}{4}r^2 + \pi R^2 \right).
	\end{equation}
\end{corollary}

Other examples can be constructed using the above approach where the drop is symmetric about a central axis, but bound to the outer portion of the tube and attached to the wall.  In these cases, the drop interfaces will be Delanay surfaces, the rotationally symmetric constant mean curvature surfaces.  We do not include an analysis of these configurations in this work.  We also note that the other possibilities in this case are asymmetric drops attached to the tube wall, and these are also outside of the scope of this work. 

\subsection{Comparison between closed-form solutions and numerically generated solutions}

In order to compare our numerical method to the energy minimizer constructed above, there are some details that differ.  The first, and primary of these differences is that the solver built by Treinen \cite{Treinen2023a} is not appropriate for constant mean curvature equations.  We are, however, able to take a small value for the gravitational constant and use this solver.  The solver worked well for $g = \texttt{1e-8}$.  We use the simplest case of  $\sigma_{01} = \sigma_{02} = \sigma_{12} = 1$.    Secondly, in the zero gravity case, the density values of the fluids are irrelevant, however in our approximation of that problem, we use our standard value of $\rho_2 = 15$ and we found the smallest errors when we picked $\rho_1 = 7.5$, giving identitical values for $\kappa_{01}$ and $\kappa_{12}$.  These differences do lead to a source of error that is not due to our method itself.

When comparing the interfaces in the two different configurations, we will use a subscript to denote the zero gravity solution.  Then we can define our absolute and relative error by
\begin{equation}
	E_{\mbox{abs}} = ||u - u_0|| + ||v - v_0|| + ||w - w_0|| + |\bar r - \bar r_0|,
\end{equation}
and
\begin{equation}
	E_{\mbox{rel}} = \frac{E_{\mbox{abs}}}{||u_0|| + ||v_0|| + ||w_0|| + |\bar r_0|},
\end{equation}
where $||\cdot||$ denotes the 2-norm applied using the Chebyshev points generated by the solver where we simply evaluate our explicit formulas from the zero gravity solution at the same points.

These errors when comparing the wall bound floating drops are $E_{\mbox{abs}} = \texttt{3.6738e-07}$ and $E_{\mbox{rel}} = \texttt{7.5411e-08}$.  We note that there is a change in about one order of magnitude when we change the gravitational constant by an order of magnitude.

\section{Conclusions}
\label{sec:conclusions}

We have considered floating drops that can be described by configurations based on ordinary differential equations.  These configurations lead to centrally located drops and symmetric wall-bound drops in $\mathbb{R}^3$ and additionally when we consider the lower-dimensional problem, we are able to find asymmetric wall-bound drops with either the entire drop area on one wall, or unevenly split drops adjacent to both walls.  In all of these situations, for any given set of physical parameters we are able to find the two distinct floating drops in $\mathbb{R}^3$ and three distinct floating drops in $\mathbb{R}^2$ in addition to the one-parameter family of a volume (area) of fluid split between two drops adjacent to each wall.  The solutions of the free boundary problems for floating drops exhibit no uniqueness at all.

We then explored the potential energy of these different configurations, and we reported on cases where the centrally located drop was the energy minimizer, the symmetric wall-bound drop was the energy minimizer, and the asymmetric wall-bound drop was the minimizer in $\mathbb{R}^2$, giving symmetry breaking.  In no cases we explored in $\mathbb{R}^2$ did we find the asymmetric split-drop minimized the energy.  We did find many cases where the (presumed) energy minimizer was non-unique.  While we do not classify all possible configurations, the examples in $\mathbb{R}^2$ are likely to exhaust reasonable competitive options.  In $\mathbb{R}^3$, the asymmetric solutions would involve systems of nonlinear elliptic PDEs, and the numerical methods to treat these free-boundary problems are outside of the scope of this work.  It would be interesting to revisit these examples if a robust accurate numerical method was developed that could treat some of the subtleties of the capillary equations in more general domains.

We gave guiding heuristics for when a centrally located drop is likely to be the energy minimizer, and we also gave an example of when this is false.  We gave an example of when the symmetric split-wall drop is the energy minimizer, and we gave evidence for our findings that the asymmetrical split-wall drop is never the energy minimizer in $\mathbb{R}^2$.

In our explorations of this nine-dimensional parameter space, we found four types of energy parameters, those being the bulk parameters of volume (area) and tube radius, density values, surface tensions, and wetting parameters.  We gave examples of parameter studies in each of these families.  Specifically, we also gave a framework for a volume parameter study that could lead to verification with physical experiments.  We observed that some symmetry in certain parameters lead to a reflection symmetry in the drop profile.  We gave some indication to how the parameter space could be further explored by looking at pairs of parameters and classifying the number of equal energy cases found. We finished our study with rigorous examples of the non-uniqueness of soluions to the free boundary problem and then we used those examples to validate our numerical method.

As we stated earlier in the paper, we make no claim to have exhausted all of the interesting cases in this somewhat large parameter space.  We have merely given a glimpse into the complexity of these problems with our numerical techniques.


\end{document}